\documentclass[journal]{IEEEtran}
\usepackage{xcolor}
\usepackage{verbatim}
\usepackage{multirow}
\usepackage{cite}
\usepackage{amsmath,amssymb,amsfonts, mathtools}
\usepackage{algorithm}
\usepackage{algpseudocode}
\usepackage{graphicx}
\usepackage{textcomp}
\usepackage{tabularx}
\usepackage{array}
\usepackage{bbm}
\usepackage[normalem]{ulem}
\usepackage{amsthm}
\theoremstyle{definition}
\newtheorem{definition}{Definition}

\usepackage{lipsum}
\usepackage{cuted}
\usepackage{color,soul}

\def\BibTeX{{\rm B\kern-.05em{\sc i\kern-.025em b}\kern-.08em
    T\kern-.1667em\lower.7ex\hbox{E}\kern-.125emX}}

\usepackage{footnote}
\makesavenoteenv{tabular}
\makesavenoteenv{table} 

\begin{document}

\title{Multi-Agent Context Learning Strategy for Interference-Aware Beam Allocation in mmWave Vehicular Communications}

\author{\IEEEauthorblockN{Abdulkadir Kose\IEEEauthorrefmark{1},
Haeyoung Lee\IEEEauthorrefmark{2}, Chuan Heng Foh\IEEEauthorrefmark{3}, Mohammad Shojafar\IEEEauthorrefmark{3}}\\
\IEEEauthorblockA{\IEEEauthorrefmark{1}Department of Computer Engineering, Abdullah Gul University, Kayseri, TR\\
\IEEEauthorrefmark{2}{School of Physics, Engineering and Computer Science, University of Hertfordshire, Hatfield, UK}\\
\IEEEauthorrefmark{3}5GIC \& 6GIC, Institute for Communication Systems (ICS),
University of Surrey,
Guildford, UK\\
Email: abdulkadir.kose@agu.edu.tr, h.lee@herts.ac.uk, \{c.foh, m.shojafar\}@surrey.ac.uk}}

\iffalse
\author{Michael~Shell,~\IEEEmembership{Member,~IEEE,}
        John~Doe,~\IEEEmembership{Fellow,~OSA,}
        and~Jane~Doe,~\IEEEmembership{Life~Fellow,~IEEE}% <-this % stops a space
\thanks{M. Shell was with the Department
of Electrical and Computer Engineering, Georgia Institute of Technology, Atlanta,
GA, 30332 USA e-mail: (see http://www.michaelshell.org/contact.html).}% <-this % stops a space
\thanks{J. Doe and J. Doe are with Anonymous University.}% <-this % stops a space
\thanks{Manuscript received April 19, 2005; revised August 26, 2015.}}
\fi

% The paper headers

\iffalse
\markboth{Journal of \LaTeX\ Class Files,~Vol.~14, No.~8, August~2015}%
{Shell \MakeLowercase{\textit{et al.}}: Bare Demo of IEEEtran.cls for IEEE Journals}
\fi
% The only time the second header will appear is for the odd numbered pages
% after the title page when using the twoside option.
% 
% *** Note that you probably will NOT want to include the author's ***
% *** name in the headers of peer review papers.                   ***
% You can use \ifCLASSOPTIONpeerreview for conditional compilation here if
% you desire.

% If you want to put a publisher's ID mark on the page you can do it like
% this:
%\IEEEpubid{0000--0000/00\$00.00~\copyright~2015 IEEE}
% Remember, if you use this you must call \IEEEpubidadjcol in the second
% column for its text to clear the IEEEpubid mark.

% use for special paper notices
%\IEEEspecialpapernotice{(Invited Paper)}

\ifpdf
    \graphicspath{{FIG/PNG/}{FIG/PDF/}{FIG/}}
\else
    \graphicspath{{FIG/EPS/}{FIG/}}
\fi

% make the title area
\maketitle

% As a general rule, do not put math, special symbols or citations
% in the abstract or keywords.
\begin{abstract}

Millimeter wave (mmWave) has been recognized as one of key technologies for 5G and beyond networks due to its potential to enhance channel bandwidth and network capacity. 
The use of mmWave for various applications including vehicular communications has been extensively discussed. However, applying mmWave to vehicular communications
faces challenges of high mobility nodes and narrow coverage along the mmWave beams. Due to high mobility in dense networks, overlapping beams can cause strong interference which leads to performance degradation. As a remedy, beam switching capability in mmWave can be utilized. 
Then, 
frequent beam switching and cell change become inevitable to manage interference, which increase computational and signalling complexity. 
In order to deal with the complexity in interference control, we develop a new strategy 
%HL for Contextual Bandit 
called Multi-Agent Context Learning (MACOL)%HL
, which utilizes Contextual Bandit 
to manage interference while allocating mmWave beams to serve vehicles in the network. 
Our approach demonstrates that by leveraging knowledge of neighbouring beam status, 
%HL We show that with the knowledge of the status of the neighbouring beams, 
the machine learning agent can 
%HL is able to 
identify and avoid potential interfering transmissions to other ongoing transmissions.
Furthermore, we show that even %at HL  
under
heavy traffic loads, our 
proposed MACOL 
strategy 
is able to maintain  
low interference levels 
at % HL 
around 10\%.

\end{abstract}

% Note that keywords are not normally used for peerreview papers.
\begin{IEEEkeywords}
Vehicular networks, mmWave, beam management, machine learning, multi-armed bandit. 
%HL IEEE, IEEEtran, journal, \LaTeX, paper, template.
\end{IEEEkeywords}

\IEEEpeerreviewmaketitle

\section{Introduction}

%--------HIGH-LEVEL INTRO ---------

\IEEEPARstart{T}{he} trends towards connected and autonomous vehicles (CAVs) have 
gained significant momentum in 
both industry and academia,
%HL due to benefits it offers such as less traffic congestion, safer driving and less
driven by the numerous benefits they offer, such as reduced traffic congestion, enhanced driving safety, and decreased CO2 emission~\cite{suthaputchakun2012applications,5GAmericasWhitePaper.2018}. 
{\color{black} Traditional autonomous driving research focused on a vehicle's independent perception using onboard sensors like radar, lidar, and camera~\cite{parekh2022review}. However, these sensors have limitations in field of vision, radar range and object recognition~\cite{bagheri20215g,fu2021network}. To overcome these, there is a growing interest in leveraging connected intelligence and communication networks, with Vehicle-to-Everything (V2X) communication, especially 5G NR-based cellular V2X, emerging as a crucial solution, offering high-speed, low-latency, and reliable wireless links~\cite{rasheed2020intelligent,sakaguchi2021towards}.}

{\color{black} 
V2X facilitates the proactive exchange of vital information among different entities including networks and vehicles, resulting in an overall improvement in driving quality~\cite{lee2021design}. This capability significantly enhances a multitude of applications, including platooning, cooperative adaptive cruise control (CACC), lane merging and other intelligent transportation systems (ITS)s, thereby optimizing the traffic flow and operational efficiency \cite{V2XuseCases}. Notably, a study in \cite{cao2017testbed} demonstrated the efficacy of 5G-V2X communications in successfully activating emergency brakes with a response time of under one millisecond, thereby improving safety. Furthermore, V2X aids CAVs in safely passing intersections, reducing congestion \cite{fu2021network}.}

{\color{black}In the realm of V2X applications, encompassing cooperative automated driving and enhanced infotainment services, the need for high data rates is crucial, particularly for the exchange of sensory image data, real-time traffic updates, and efficient path planning. For example, for safe and efficient driving, having up-to-date high definition (HD) 3D map would be useful \cite{HD3DMap,TR26.985}. However, this map can quickly become outdated due to factors such as road construction. In response, vehicles can leverage V2X communication with high data rates to download the latest HD 3D map data on-demand from the infrastructure. This approach ensures keeping HD 3D map data updated, and allows it to promptly reflect changes on shorter time scales. Moreover, scenarios may arise where vehicles upload onboard sensor data to cloud and/or edge servers. These servers can then utilize the data for constructing crowdsourced HD 3D map information and performing traffic analysis. To fulfill the demanding requirements for high data rates in these diverse applications, millimeter-wave (mmWave) technology has merged as a promising candidate for 5G and beyond networks \cite{Rahim22Svy}, particularly within the context of vehicular applications~\cite{5gMaz21,zhao2018vehicular, sakaguchi2021towards, rasheed2020intelligent}.}

Despite the benefits of mmWave access, transmissions in %HL at 
such high frequency range %HLies 
(above 6GHz) are more susceptible to blockages and often suffer from significant  propagation loss which can hinder communication reliability \cite{Rahim22Svy}. Dense deployment of mmWave cells and directional transmission are two measures taken to counter 
non-line of sight (NLOS) cases caused by 
%HL due to 
blockages and high propagation loss. 
%The directional transmission in mmWave, often called a mmWave beam, is obtained when mmWave transmitters focus its transmission power towards a specific direction to provide a practical transmission range. 
However, these measures introduce %HL Such directional transmission and densification, however, introduces 
new challenges for the cell level and beam level mobility management in vehicular networks due to dynamic topology. The narrow and short coverage of mmWave beams along with fast moving vehicles can result in %HLlead to 
a short period of sojourn time for a vehicle within a beam which in turns causes frequent beam selection and switching, leading to increased %HL resulting high
overhead\cite{Kose20PIMRC}. 
For the connectivity establishment, there is a need for accurate beam alignment and robust connection management 
against rapid channel changes~\cite{kose2021beam}. 
Moreover, while it may seem favorable to provide users with long sojourn time connectivity, 
the longest sojourn time does 
not always guarantee the best service throughput performance throughout the transmission period in the presence of interference. 
When a user is situated in an area where the serving beam overlaps with another beam, this service period will not be interference-free, especially if both beams are simultaneously serving different users. This situation can lead to significant performance degradation. Additionally, the need for dense deployment of mmWave small cells to increase network capacity %has only escalated 
can exacerbate  
the inter-beam interference,  
%HL which diminishes 
ultimately diminishing the expected throughput gain
from 
%HL expected by 
network densification.

To capitalise the densification gain, interference-aware beam selection schemes should be considered to reduce the impact of interference. In multiple antenna-based systems, beamforming has emerged as a key technique for handling inter-cell interference. By maximizing power in  desired directions while suppressing it on interfering links, beamforming plays a crucial role in both the transmitter and receiver sides. In general, despite the potential, receiver-based beamforming aims to maximize received power based on the user feedback for the received beamforming vector and does not take into account the interference caused to the other ongoing transmissions. 
To account for the impact of interference on neighbouring users, beamforming weights can be selected to maximize the signal-to-caused interference ratio (SCIR) \cite{hassanpour2008distributed}. When multi-antenna users are considered, joint optimization of both transmitter and receiver beamforming vectors is required to achieve high SCIR. In multi-cell multi-antenna networks, however, devising effective coordination and cooperation strategies among multiple transmission points for joint decision-making on interference-aware beam transmissions remains an open research challenge~\cite{salameh2016efficient, yoon2019downlink, deng2017resource,hoshino2021study,elsayed2020machine,wang2019interference,solaija2021generalized,
%yoon2019downlink, HL-duplicated
irram2020coordinated,de2018comparing,zhang2019beam
%,wang2019interference HL-duplicated
}.
%\footnote{\color{blue}it’s unclear that beam interference management is all about interference-aware beam selection. I think this is also not generic enough. The literature here should cover both passive approach of beam interference mitigation and active approach of beam interference avoidance. Our solution falls under the second approach.\color{teal}AK:What is meant by passive and active interference mitigation? DONE} 

%\cite{salameh2016efficient}\footnote{\color{blue}What message to make here? What's the purpose? I don't see a concluding statement}.

\subsection{Related Works}

In the literature, beamforming systems are commonly considered under two catagories; switched-beam system (SBS) and adaptive array 
(AAS). In the SBS case, beams patterns are fixed with predetermined pointing directions. In AAS, special beams can be created for each user by series of antanna array processors which apply transmitter and receiver beamforming weight vectors %to communicated signals 
to adjust phase and amplitude of antenna elements. This allows the main beam lobe to be directed towards the intended user,  while sidelobes are aligned towards  interfering links. 

The following studies\cite{salameh2016efficient,yoon2019downlink}  
considered AAS.
The authors  in \cite{salameh2016efficient} proposed an efficient inter-cell interference avoidance scheme to maximize sum-rate in multi-cell antenna systems. In this system, adjacent base stations (BSs) and candidate receivers cooperatively optimize their transmitter/receiver beamforming weights. The goal is to maximize the receiver power at the served user while minimizing interference caused to users in other cells.
However, 
this work assumes that certain cells are grouped to create a virtual cell for transmission coordination for multiple users. %which coordinates the transmission of multiple users among its cells.
It %This 
requires strict time synchronization and efficient coordination among cells, which can introduce computational complexity on the BS sites. To offload computational complexity to edge and enable more efficient coordination among transmission points, the authors in \cite{yoon2019downlink} proposed a downlink interference alignment scheme 
for fog radio access networks 
where each radio remote head (RRH)
is connected to a fog access point (F-AP) via a wired link. 
To reduce inter-cell interference hence to maximize sum-rate of the network, 
F-AP decides the optimum beams for each transmission pairs upon receiving channel feedback from users in response to randomly generated beamforming matrices at each RRH. The work also emphasizes the existence of the trade-off between sum-rate performance and feedback overhead, which requires further investigations, as the feedback overhead from users to RRH increases with the number random beamforming matrices.

{Since
practical implementation of fully reconfigurable front-ends 
in AAS system are expensive
\cite{wang2019interference}}, some studies~\cite{zhang2019beam,wang2019interference, hoshino2021study} 
adopted SBS due to its low cost and less complexity. Nevertheless, frequent beam switching and cell changes are the main concerns in SBS that need to be mitigated while minimizing interference. In~\cite{zhang2019beam}, the authors proposed a lightweight beam interference suppression method that involves %by performing 
beam switching and multi-cell cooperative transmission. 
The lightweight feature is obtained through a two-fold threshold mechanism %is considered 
for user grouping, which helps mitigate the problem of frequent beam switching. % to %reduce 
%avoid frequent beam switching problem.  
For user grouping, 
interference-to-power ratio of each user is compared with 
upper and lower interference power thresholds. If interference-to-power ratio of a user is higher than the upper threshold value, the user will be added to the high interference group which requires multi-cell cooperation to suppress the interference. In case interference-to-power ratio is higher than the lower threshold value, intra-cell beam switching within the same BS is performed to avoid interference. To  address the complexity of coordination for user and beam pair for each transmission in massive antenna array using beamforming techniques, 
the work in ~\cite{wang2019interference} proposed a two-stage beam coordination approach %for a SBS, 
which groups cells into clusters to reduce intra-cluster and inter-cluster interference. 
In the intra-cluster stage, based on mutual interference level, several sectors where each sector has width of 120° and 48 beams are grouped into a same cluster. In order to manage  interference within a cluster, the coordinated scheduling is used to schedule transmissions for users at the transmission time interval (TTI)  level. On the other hand, lack of coordination between clusters leads to  interference at cell edge users. To address this issue, in the inter-cluster stage, by using time domain interference coordination approach, cluster-edge users are dynamically allocated among clusters based on interference condition to reduce inter-cluster interference. In \cite{hoshino2021study}, the authors proposed a coordinated beamforming among multiple 
BSs to reduce inter-cell interference and improve cell edge throughput performance. 
The coordinated BSs 
are connected via a beamforming control unit. Each mobile terminal measures the signal strength from each beam and feeds the received signal strength back to the serving BS. Then, each BS sends these measurements to the control unit for beamforming coordination. Coordinated beamforming is based on code-book beamforming, where beam directions are predefined, and one beam is selected for each terminal to maximize the minimum throughput of the each terminal  
and the total throughput.

Despite efforts made in  cooperative and coordinated transmission in multi-cell scenarios and optimization of the beam management to reduce interference, the above solutions have taken a passive approach which is inadequate to proactively deal with the complication in the modeling and predicting beam radiation footprint, channel conditions, and dynamics of the vehicular mobility.  
In recent years, the use of machine learning (ML) 
\cite{ML5G19} for beam management 
in mmWave systems has received a significant level of attention \cite{Ma2022ML, Yang2020ML,Yin2018ML,Ju2020ML,elsayed2020machine}.
While the conventional beam management studies including  \cite{hoshino2021study,wang2019interference,zhang2019beam,yoon2019downlink,salameh2016efficient} rely on mathematical optimization methods, these approaches 
model-based. However, accurate modelings of interference model, channel estimation, and dynamics of moving vehicle can be challenging  \cite{6gV2X}.
The use of ML 
can support the adaptive learning of the channel features, enabling reliable beam management  without relying on  accurate models  \cite{Ma2022ML}.  
Moreover,  
in high-dimensional features of the propagation scenarios that involve extensive blockage,  
a data-driven ML approach can be employed for more efficient beam selection by learning and adapting to changing environments.  

In recent studies \cite{Ju2020ML,elsayed2020machine}, the authors apply ML techniques to address the challenge of interference-aware beam allocation. Ju \emph{et al.}~\cite{Ju2020ML} design a centralized solution for user scheduling and precoding in a multi-cell mmWave network. They utilize 
deep neural network (DNN) at 
the central node to predict beamforming vector and user groups, enabling efficient assignment of cells to multiple users based on spectral efficiency considerations. The effectiveness of DNN-based  prediction is demonstrated, showcasing improvements in spectral efficiency. In another work ~\cite{elsayed2020machine}, Elsayed \emph{et al.} consider a multi-cell scenario 
employing non-orthogonal multiple access (NOMA) technology and design a ML algorithm to manage the inter-beam and inter-cell interference. Recognizing that users located within the intersection region of multiple  cells are more susceptible to  interference, the authors propose distributing a ML agent in each cell to coordinate and learn the optimal policies for  user-cell association and inter-beam power allocation to support these vulnerable users. 
This work adopts Q-learning \cite{ML5G19}, a reinforcement learning approach, and its  performance is evaluated in a scenario including two base stations and static users, demonstrating an increase in the overall sum rate. 
%{\color{red} 

Among various ML techniques,  multi-armed bandit (MAB) is considered in \cite{Sim18Trans,li2021beamselection} for beam selection in  mmWave systems. The goal of MAB learning agent is to learn the environment and apply past experience to make its decision. 
For learning, a balance between exploration and exploitation becomes crucial when there are limitations on resources such as time or number of
actions for exploration and exploitation are limited \cite{Kose2021}. 
A MAB training beam selection policy can be used to balance exploration of the set of feasible beam. 

In \cite{Sim18Trans}, the adaptive beam selection is formulated as a contextual multi-armed bandit (C-MAB) problem. The proposed online learning approach incorporates contextual information, such as the traffic pattern and the vehicle's traveling direction. A mmWave base station (BS), serving as a learning agent, autonomously learns the relationship between beam selection and data rate performance given the contextual information, including the traffic pattern and the presence of permanent or temporary blockages. Furthermore, in \cite{li2021beamselection}, the focus extends to broadcasting clustering of neighbouring vehicles that intend to download 
the same popular contents (e.g. movies), while considering their traveling directions. The  mmWave BS learns   
the appropriate beams to cover multiple vehicles requesting  the same contents within the cells, as well as the most suitable  broadcast angle along these beams. 
Since a MAB based algorithm learns the expected  performance in different contexts over time, it actually does not require a training phase and is highlighted as a fast learning algorithm.

{%\color{blue}
Despite the effort made for ML-based interference management, none of the work above has considered vehicular communication on highway scenarios. To consider mmWave vehicular communication performance on a highway scenario where BSs are deployed on both sides of the road, \cite{Tassi.01062017,giordani2018coverage} have considered stochastic geometry to analyze the performance of beam coverage and connectivity in mmWave vehicular networks. \cite{giordani2018coverage} focuses on beam alignment to enhance connectivity and data rate. The authors demonstrated that the performance of vehicle communication performance relies heavily on the periodicity of beam alignment, vehicle speed, beamwidth, and base station density. It is also stated that the stability of connections relies on both beam alignment and sufficient signal quality. \cite{Tassi.01062017} focuses on the modeling SINR outage probability and rate coverage probability. It is found that SINR outage probability increases with BS density increases in sparse networks and large beamwidths as interference becomes apparent. However, these works have not proposed an ML-based solution to improve network performance.
}

\subsection{Our Contribution}

Cooperation among multiple cells to mitigate interference requires various information exchange in order to collaboratively decide which beams from which cell should be used to allocate to which users. Considering the fast changing of network topology due to vehicle mobility, %{\color{blue}\sout{this may cause additional signalling overhead and delay (e.g. high Xn interface delay), where instant decision is required} 
keeping up with the topology changes requires low latency message exchanges among BSs for prompt decision making. %{\color{blue}\sout{Instead, we consider a central node \textcolor{red}{(CN)} to collect status of all beams, i.e. on/off and distribute them instantly whenever needed by any BS. In this regard we proposed a lightweight and fast intelligent beam allocation method.} 
To deal with the fast-changing environment, we propose a ML algorithm to adaptively respond to the changing environment without significant signalling overheads. Our design uses a central node (CN) to collect operating status of all beams and provide the status information whenever needed by any BS. {%\color{blue}
In a typical mobile network setup, a set of BSs are clustered and managed by a central unit, the central unit can be served as the CN. As the collection only occurs when any BS changes its status between idle and active which does not happen frequently, the signaling overhead is low.
} The proposed method employs an edge-assisted contextual %Multi-armed bandit 
MAB to reduce inter-cell and inter-beam interference utilising the beam operating status of the neighbouring cells as a context. In this regards, each BS makes decision independently in a distributed manner with the context shared by the CN. In our design, the CN does not control the beam transmissions directly, rather its role is simply pulling and pushing information between BSs without making any decision in the process. This role can be easily fulfilled by existing radio access network (RAN) architecture where the information can be delivered promptly using existing control channels.
%\sout{rather it collects and disseminates the stored status of the all beams in the region of interest to bypass long delays and signalling overhead for cooperation and beam-switching. Note that each BS is directly connected to CN via optical fiber so delay becomes insignificant at short distance.}} 
%
%This setup is also an advantage in terms of communication cost in a  distributed multi-agent MAB scenario where each agent has to exchange messages with other nodes \cite{hillel2013distributed,agarwal2021multi}.
%\footnote{\color{blue}I don't understand the logic of this statement. -HL:Done- Kadir: plz check? REMOVED}
%
In this paper, we apply our above mentioned design to a highway scenario with densely deployed multi-cell mmWave BSs along the highway to provide high data rate transmissions to the passing vehicles.
The following describes the main contributions of this paper.
\begin{itemize}
  \item We develop an analytical model which derives the upper-bound for vehicle sojourn time in a beam of a dense mmWave network. Specifically, we formulate directly the vehicle displacement within a beam. The derived upper-bound results are also used to benchmark the performance of a particular setup.
  \item We also study the impact of the interference on vehicle service period withing a beam. By using geometrical analysis, we derived the probability of a vehicle being interfered by other beams which it also resides. 
  \item We propose a lightweight, fast and intelligent distributed beam allocation ML solution. The proposed ML algorithm called \emph{Multi-Agent Context Learning} (MACOL) algorithm %{\color{blue}\sout{employ the contextual Multi-armed bandit to reduce inter-cell (inter-beam) interference utilising the state information about each beam of the neighbouring cells as context.} \hl{
  extends MAB by introducing context learning in the process to enable the ML agent to learn potential transmission interference based on the shared contexts among distributed ML agents.
%HL  \item We also provide regret analysis ....
\end{itemize}

The remainder of the paper is organized as follows. Section \ref{sec:analytical} presents the modeling and derivation of beam service period which is used to study the impact of interference in a densely deployed mmWave network. In Section \ref{sec:C-MAB}, the proposed interference-aware beam allocation algorithm based on contextual MAB is introduced. The performance validation are given in Section~\ref{sec:result} showing the effectiveness of our proposed algorithm with key findings and shares potential directions for future research. Finally, the conclusions
are given in Section~\ref{sec:conclusion}. 
%\footnote{\color{blue}Missing Section IV --> HL: done}

\section{Analysis of Vehicle Service Period within a Beam}
\label{sec:analytical}

\begin{table}[]
\caption{Notation Table}
\label{tbl:notation}
\begin{tabular}{|>{\centering}m{0.08\textwidth}|p{0.35\textwidth}|}
\hline
Notations & Descriptions \\ \hline
$B_N$ & The total number of beams in the scenario. \\ \hline
$b_i$ & The $i$-th beam in the scenario, where $i\in \mathcal{B}_N$ and $\mathcal{B_N}=\{1,2,...,B_N\}$. \\ \hline
$(x_i,y_i,\theta_i)$ & A 3-tuple describing the BS location and pointing direction for beam $b_i$. \\ \hline
$\mathcal{B_L}$ & The layout of the deployment in the scenario. It is a set of $(x_i,y_i,\theta_i)$ tuples,  $\forall i\in \mathcal{B_N}$. \\ \hline
$R_B$ & The radius of a beam. \\ \hline
$\Omega_B$ & The beamwidth of a beam. \\ \hline
$|D_i|$ & In a highway scenario, it is the distance between the BS operating beam $b_i$ and the edge of the highway. \\ \hline
$r_i\angle{\phi_i}$ & A location relative to beam $b_i$ where $r_i$ is the distance from the beam radiation origin and $\phi_i$ is the angle from the beam pointing angle. \\ \hline
$r_{k|i}\angle{\phi_{k|i}}$ & A location relative to beam $b_k$ translated from $r_i\angle{\phi_i}$. That is, the location $r_i\angle{\phi_i}$ is re-written as $r_{k|i}\angle{\phi_{k|i}}$ which is relative to beam $b_k$. \\ \hline
$A_B$ & The area of a beam, $A_B=\frac{1}{2}R_B^2 \Omega_B$ \\ \hline
$A_I(k)$ & The area of beam $b_k$ overlapped with other beams given the layout $\mathcal{B_L}$. \\ \hline
$p_i$ & The probability that beam $b_i$ is active at an arbitrary observation time. \\ \hline
$\tilde{L}$ & The random variable describing the vehicle travelling distance in a beam. \\ \hline
$F_L(l)$ & The Cumulative Distribution Function (CDF) of $\tilde{L}$. We further modify $F_L(l)$ to $F_L^*(l)$ for highway scenario, and $\breve{F}_L^*(l)$ to include interference consideration for highway scenario.\\ \hline
$J_k(\cdot)$ & The probability that a vehicle within beam $b_k$ starting at $(r_k,\phi_k$) departs the beam after travelling $l$ distance. We also modify $J_k(\cdot)$ for several considerations: (i) $J_k^*(\cdot)$ for highway scenario, (ii) $\breve{J}_k^*(\cdot)$ to also include interference consideration for highway scenario. \\ \hline
$\vec{S}(t)$ & The context represents the environment state at time $t$. ML agent $i$ then derives its own context $\vec{S}_i(t)$ using masking. \\ \hline
$\mathcal{C}_i$ & The set containing all contexts observed by ML agent $i$. \\ \hline
$\rho_i(t)$ & The reward computed by ML agent $i$ for the connection established at time $t$. \\ \hline
%HL ??? & The reward... \\ \hline
\end{tabular}
\end{table}

In this section, we perform beam analysis
%\footnote{Our derivations are rigorously validated. We share unit tests we performed with full source code available at https://github.com/cfoh/beam-analysis.} 
where we derive the \emph{service period} of a vehicle while passing a beam. We define the vehicle service period within a beam to be the time duration that the vehicle is served by a beam from when the vehicle has connected to the beam to when the vehicle has lost the connection. The connection and disconnection are triggered by a handover event. A handover event occurs when a vehicle is about to move out of the coverage of its current beam connection. When this happens, the vehicle seeks for a target beam to handover to. Considering random beam deployment and a target beam is found, the location of the vehicle in its target beam follows a Poisson Point Process (PPP) uniformly distributed within the target beam. Each beam, say $b_i$ in a deployment scenario can be characterized by its location of radiating origin and direction $(x_i,y_i,\theta_i)$, its beam sector radius $R_B$ and beamwidth setting $\Omega_B$.

{%\color{blue}
\subsection{Channel Model}

We consider 3GPP Band n257 channel for the mmWave transmission. The bandwidth is set to 50 MHz. For the analysis work, we use geometric framework for our modeling approach~\cite{Semiari2018,Li2016,tatino2017beam}. With the directional antenna feature of a mmWave BS, we approximate the beam coverage to have a shape of a sector, and its coverage follows the definition given below.
} % color{blue}
\begin{definition}[Beam Coverage for Geometric Framework]
In geometric framework, the coverage of a beam is defined by a particular geometric shape. In this paper, we assume that the shape of the footprint of a beam is a sector. A vehicle can communicate with the base station radiating the beam only if it is located inside the footprint of the beam.
\end{definition}

{%\color{blue}

In practice, the range of a beam transmission depends on the channel pathloss and other settings related to antennas and transceivers. In the simulation, we also consider beam coverage for practical setup to show the effectiveness of our proposed ML algorithm. In this setup, following pathloss for mmWave channel \cite{Sun16}
\begin{equation}
\label{eqn:pathloss}
PL(d) = 32.4 + 20 \log(f_c) + 10 \alpha \log(d)
\end{equation}
where $\alpha$ is 2.1 when $d<54$ and 3.4 otherwise. Other settings are given in Table~\ref{tab:txParams}. In this setup, while the beamwidth setting follows that of geometric framework, the beam radius $R_B$ is determined by the received signal strength. The beam radius is the maximum range where the Signal-to-Noise Ratio (SNR) of the received transmission exceeds a predefined threshold. 

} % color{blue}

\begin{table}[tbp]%\color{blue}
\caption{Transmission parameters setting}
\begin{tabular}{  m{3.4cm} | m{4.6cm}  } \hline
Parameter & Value  \\ \hline \hline
Central frequency, $f_c$ & 28GHz \\ \hline
Bandwidth & 50MHz \\ \hline
Transmit power & 20dBm \\ \hline
SNR threshold & -5dB \\ \hline
Beamforming gain & 9dB \\ \hline
Noise figure & 7dB \\ \hline
\end{tabular}
\label{tab:txParams}
\end{table}

\begin{figure}[!tp]
\centerline{ \begin{tabular}{c}
\includegraphics[width=0.31\textwidth,height=!]{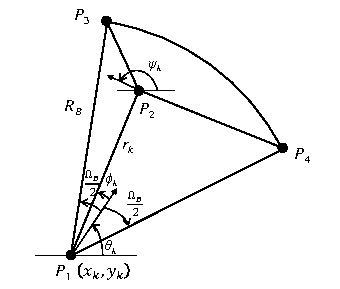} \\
\end{tabular}}
  \caption{Illustration of a typical beam sector geometry layout with a BS at $P_1$ radiating a beam pointing at $\theta_k$, and a vehicle located at $P_2$ travelling in the direction $\psi_k$ within the beam. From the perspective of the beam, the vehicle is location at $r_k\angle{\phi_k}$ relative to the beam.}
  \label{fig:beamGeom}
\end{figure}

Fig.~\ref{fig:beamGeom} illustrates a typical beam sector $b_k$ radiating from the point $(x_k,y_k)$ towards the direction $\theta_k$ with radiation range of $R_B$ and beamwidth of $\Omega_B$. The figure also shows a vehicle located at $P_2$ within beam $b_k$ travelling in the direction $\psi_k$. We specify the location of the vehicle as $r_k\angle{\phi_k}$ which is relative to beam $b_k$. Precisely, the vehicle is located $r_k$ away from the origin of beam $b_k$ with an angle $\phi_k$ from the beam pointing angle. The vehicle is said to reside within beam $b_k$ if $r_k \in [0,R_B]$ and $\phi_k \in [-\frac{\Omega_B}{2},\frac{\Omega_B}{2}]$.
%
%..., a handover event may occur at any random location uniformly distributed in $A$. At this location, the vehicle may detect a number of beams from its nearby BSs for the handover, and its location within each beam is also random with uniform distribution. We assume that the vehicle continues to maintain the same direction moving constantly after the handover. Given the directional antenna feature of mmWave BS, the considered beam coverage is assumed to have a shape of a sector.

%\subsection{Generic Scenario}
%
%We consider a beam $b$ defined by its direction $\theta_{b}$, beam width angle $\theta_{s}$ and radius $R$, with a coverage area $A_b$. At the time of handover decision, the beam is visible by a vehicle $v$ at a location with polar coordinate $V(r,\theta_{v})$ moving in a particular direction. Assuming that the vehicle has handed over to this beam and it continues to move in the same direction at $\phi$, the vehicle displacement within this beam can be determined by the distance between the current location of the vehicle and a point at the edge of the beam where the vehicle departs some time later. The probability that the vehicle displacement is shorter than a particular value $l$ can be determined by
% $P\left((r,\theta_{v}) \right)$
%
We consider that a vehicle $v$ is about to lose its current connection and has decided to perform a handover to a target beam $b_k$. At the time of the handover, vehicle $v$ is located at $r_k\angle{\phi_k}$ relative to beam $b_k$ moving at a particular direction with a constant speed. Assuming that the vehicle has handed over to this beam and it continues to move in the same direction with the same speed, the vehicle service time period depends on its travelling distance within the beam, and the distance is determined by the length between the current location of the vehicle and a point at the edge of the beam where the vehicle departs some time later. During this period, the vehicle receives connection service from the beam, thus the period is the vehicle service period.

Without loss of generality, we consider a north pointing beam in our analysis as shown in Fig.~\ref{fig:beamGeom}. In other words, we apply condition where $0\leq\theta_k\leq\pi$ in our analysis. For south pointing beams, we can simply apply a $180^\text{o}$ rotation transformation about the location of beam $b_k$ and reuse the same derivation approach for the analysis.

Let $\tilde{L}$ be the random variable describing the vehicle travelling distance within the beam. The cumulative distribution function (CDF) of $\tilde{L}$, which is the probability that the vehicle travelling distance in beam $b_k$ is shorter than a particular value $l$ can be determined by  
%\textcolor{red}{HL: $F_L(l) --> {F_\tilde{L}}(l)$?}\footnote{\color{blue}Nope. $L$ is the base variable, $\tilde{L}$ is its corresponding r.v., $F_L(l)$ is its CDF. If needed, $f_L(l)$ will be its pdf.}

%
%\begin{equation}\label{eqn:cdfL}
%  \mathbb{P}(L \leq l) = \frac{1}{A_{b}} \int_{0}^R \int_{\Theta_{1}}^{\Theta_{2} } J(r,\theta) R d\theta dr
%\end{equation}
\begin{equation}\label{eqn:cdfL}
F_L(l) = \frac{1}{A_{B}} \int_{0}^{R_B} \int_{\Theta_{1}}^{\Theta_{2}} r_k J_k(r_k,\phi_k,l) d\phi_k dr_k
  \end{equation}
where $\Theta_{1} = -\frac{\Omega_{B}}{2}$, $\Theta_{2} = \frac{\Omega_{B}}{2}$, $A_{B}$ is the area of the sector, and $J_k(r_k,\phi_k,l)$ is the probability that the travelling distance of a vehicle in beam $b_k$ is shorter than $l$. In other words, it is the probability that a vehicle starting at location $r_k\angle{\phi_k}$ relative to beam $b_k$ will depart the beam after travelling $l$ distance. The derivation of $J_k(r_k,\phi_k,l)$ is given in Appendix~\ref{appendix:Jk}. Finally, knowing the travelling distance, the vehicle service period can be readily obtained with the vehicle speed.

%$\mathcal{I_{L}}(\cdot)$ is an indicator function defined as
%
%\begin{equation}\label{indFunc}
%\mathcal{I_{L}}(\hat{l}_{r,\theta_{v},\phi})=
%  \begin{cases}
%    1       & \quad \text{if } \hat{l}_{r,\phi,\psi} \leq l, \\
%    0        & \quad  \text{otherwise}.
%  \end{cases}
%\end{equation}
%
%The detail derivation of the condition that $\hat{l}_{r,\phi,\psi}\leq l$ is given in Appendix~\ref{appendix:beamL}

\subsection{Highway Scenario}

In the above, we have presented the formulation of a vehicle travelling distance within its target beam. While this setup is valid for mobility scenario in an urban region in general, the setup for a highway scenario is slightly different. For highway scenario, BSs are usually deployed on the road side with a distance away from the highway. Vehicles mobility on the highway is limited to two directions. The results developed in the previous subsection requires further treatment.

\begin{figure}[!pt]
    \centering
    \includegraphics[width=0.38\textwidth,height=!]{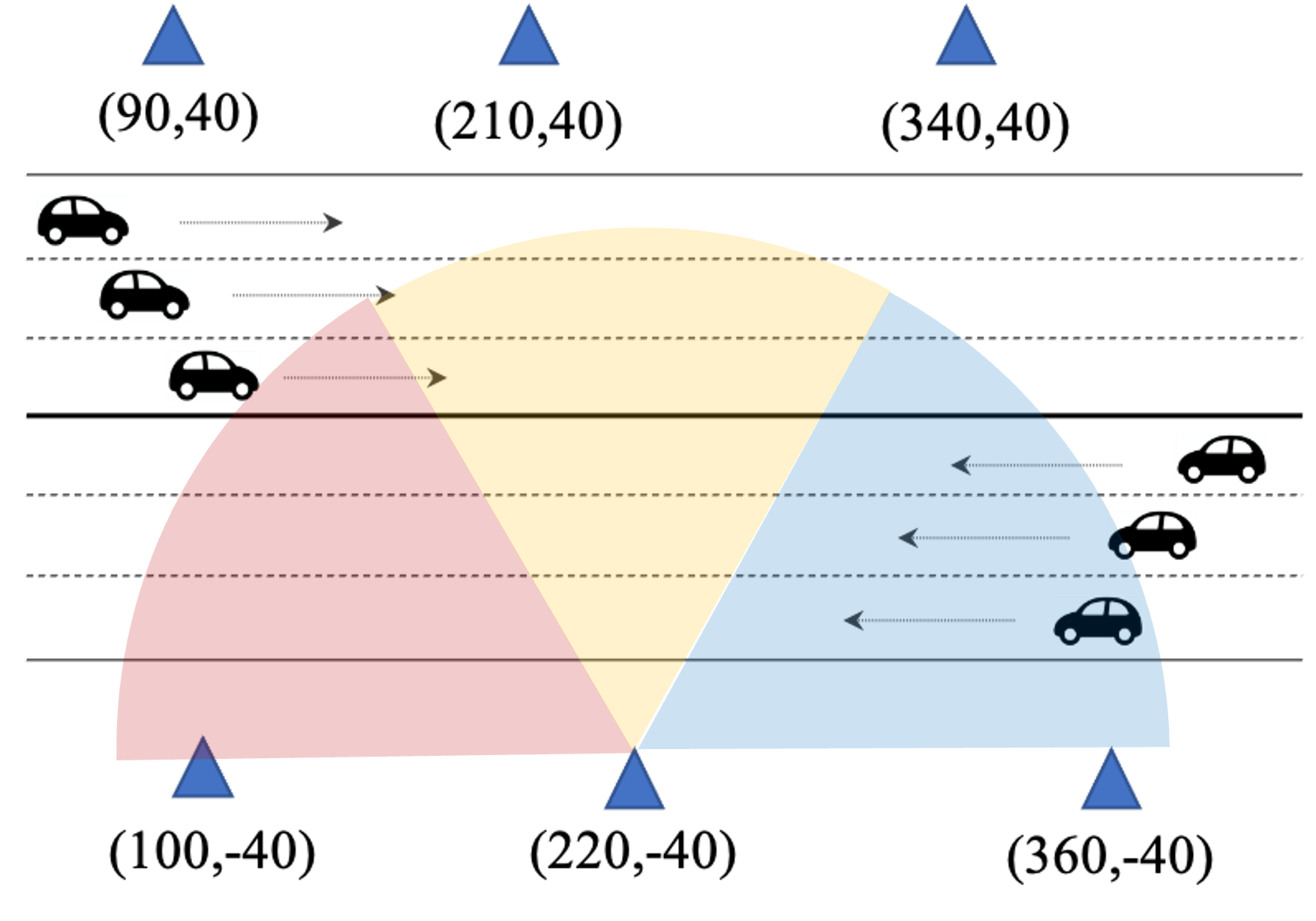}%{FIG/highway.png}
    \caption{Illustration of a highway layout with positions of BSs. 
    The illustration shows the center BS situated at the south side of the highway radiating three north-pointing beams each with beamwidth of $60^{\circ}$}% and range of 80 m.} % HL 50 m.}   
    \label{fig:roadlayout}
\end{figure}

Fig.~\ref{fig:roadlayout} shows a scenario of dense small cell BSs deployed along a highway. We assume that there are a number of BSs deployed along a section of a straight highway on both sides, which are grouped into a cluster. The highway section can appear in any orientation on a map, for the purpose of explanation, we normalize the orientation to consider that the highway section runs horizontally along the x-axis in the east-west direction. The location of each BS is shown in the figure.

In our scenario, each BS consists of 6 sectors with and each sector radiates an mmWave beam with a beamwidth of $60^o$. Out of the 6 beams from each BS, only 3 beams cover the highway. In the scenario, 3 BSs are located on the north and 3 on the south side of the highway. For the BSs located on the south (resp. north) side of the highway, only the beam pointing north, north-east, north-west (resp. south, south-east, south-west) are serving the highway. As a result, the highway is served by 18 beams.

Let $(x_k,y_k)$ and $\theta_k$ be the location and pointing direction of beam $b_k$, respectively. The beam is deployed with a distance of $D_k$ away from the edge of the highway, and the width of the highway is $H$. Since all vehicles must travel on the highway, \eqref{eqn:cdfL} is revised as follows to limit the vehicle starting location to be on the highway
\begin{equation}\label{eqn:Fstar}
F_L^*(l) = \frac{\int_{\Theta_{1}}^{\Theta_{2}} \int_{R_n}^{R_f} r_k J_k^*(r_k,\phi_k,l,\psi_k) dr_k d\phi_k}{\int_{\Theta_{1}}^{\Theta_{2}} \int_{R_n}^{R_f} r_k dr_k d\phi_k} 
\end{equation}
with
\begin{equation}
\begin{array}{c}
R_n = \min\left\{\frac{D_k}{\sin(\theta_k+\phi_k)},R_B\right\} \\
R_f = \min\left\{\frac{D_k+H}{\sin(\theta_k+\phi_k)},R_B\right\}
\end{array}
\end{equation}
where $R_n$ and $R_f$ are the nearest and farther valid distances to bound locations of vehicles within the highway. The quantity $J_k^*(r_k,\phi_k,l,\psi_k)$ is the probability that the travelling distance of a vehicle starting at location $r_k\angle{\phi_k}$ relative to beam $b_k$ travelling towards $\psi_k$ direction within beam $b_k$ is shorter than $l$, and thus the vehicle will depart the beam after moving $l$ distance. The derivation of $J_k^*(r_k,\phi_k,l,\psi_k)$ is given in Appendix~\ref{appendix:Jstar}.

%Equation~\eqref{eqn:Fstar} considers that when a beam is available to serve a vehicle, it unbiasedly selects a vehicle within its coverage. Thus it presents the average service period where serving vehicles start at any location within the beam. We can further derive the upper bound of this measure by considering an optimal vehicle selection where for a specific lane, the beam can somehow choose the vehicle just entering the beam coverage.
%
%Given a particular deployment layout $\mathcal{B_L}$, for each beam $b_k$, we can determine the number of lanes covered by the beam and the start location of each lane. Let $\mathcal{L}_k=\{(r_{s_1},\phi_{s_1}),(r_{s_2},\phi_{s_2}),...\}$ be a set collecting the start locations of all lanes, replacing the random locations in \eqref{eqn:Fstar} by the locations in $\mathcal{L}$, we derive the upper bound of \eqref{eqn:Fstar} to be
%
%\begin{equation}\label{eqn:Fupper}
%\hat{F}_L^*(l) = \frac{\sum_{(r_k,\phi_k)\in \mathcal{L}_k} %J_k^*(r_k,\phi_k,l,\psi_k)}{|\mathcal{L}_k|}.
%\end{equation}

\subsection{Impact of Interference}

We further study the impact of interference on the vehicle service period. Particularly, we investigate the reduced vehicle service period which is the service period when the connection does not experience interference, i.e. interference-free. Based on geometric framework, the interference condition is defined as follows.
\begin{definition}[Interference for Geometric Framework]
In geometric framework, a vehicle currently served by a beam is said to experience interference if it is also located within the coverage of another beam that is serving another vehicle at the same time. 
\end{definition}

{%\color{blue}
The above definition corresponds to a directional transmitter and an onmi-directional receiver. The definition of interference also represents the pessimistic performance. In practice, a transmission can tolerate some level of interference. A transmission being interfered by another transmission will suffer from degraded SINR, but if the received SINR is higher than the SNR threshold, the transmission may survive. In the simulation, we also consider interference for practical setup. We shall show the impact of interference on the performance for this setup in simulation. For the following analysis, we focus on the interference for geometric framework.
} % \color{blue}

Let beam $b_k$ be the beam sector serving a moving vehicle, and another beam $b_i, i \neq k$ be the interfering beams. Let $r_k\angle{\phi_k}$ be the vehicle location relative and residing in beam $b_k$.
%where $r_k \in [0,R_B]$ is the distance from the beam sector origin and $\phi_k \in [-\frac{\Omega_B}{2},\frac{\Omega_B}{2}]$ is the angle of the location away from the beam pointing angle. 
While the location $r_k\angle{\phi_k}$ is expressed relative to beam $b_k$, we can translate and rewrite its location as $r_{i|k}\angle{\phi_{i|k}}$ which is relative to another beam $b_i$. We first find its absolute Cartesian location $(\hat{x}_k,\hat{y}_k)$ by
\begin{equation}
\begin{array}{l}
\hat{x}_k = x_k + r_k \cos(\theta_k + \phi_k) \\
\hat{y}_k = y_k + r_k \sin(\theta_k + \phi_k) \\
\end{array} 
\end{equation}
then we translate the above to the location expression $r_{i|k}\angle{\phi_{i|k}}$ relative to beam $b_i$ by
\begin{equation}
\begin{array}{l}
r_{i|k} = \sqrt{(x_i-\hat{x}_k)^2 + (y_i-\hat{y}_k)^2} \\
\phi_{i|k} = \tan^{-1}\frac{\hat{y}_k-y_k}{\hat{x}_k-x_k} - \theta_i \\
\end{array} 
\end{equation}
where $\phi_{i|k}$ is conditioned to ensure that $|\phi_{i|k}-\theta_i| \le 2\pi$ if needed. Given the above results, we can seamlessly translate the location $r_k\angle{\phi_k}$ relatively to beam $b_k$ to $r_{i|k}\angle{\phi_{i|k}}$ which is relative to beam $b_i$.

The vehicle experiences interference from other beams if it is also resided in other beam sectors. We can determine whether a vehicle at $r_k\angle{\phi_k}$ within beam $b_k$ experiences no interference from another beam $b_i$ by translating its location to $r_{i|k}\angle{\phi_{i|k}}$ and then determine using the following calculation
\begin{equation}
\mathcal{P_I}(r_{i|k},\phi_{i|k},p_i) = \left\{
\begin{array}{ll}
      1-p_i, & r_{i|k}\leq R_B, -\frac{\Omega_B}{2}\leq\phi_{i|k}\leq \frac{\Omega_B}{2} \\
      1, & \text{otherwise.} \\
\end{array} 
\right.
\end{equation}
In the above, $p_i$ is the probability that beam $b_i$ is active at an arbitrary observation time, and $\mathcal{P_I}(r_{i|k},\phi_{i|k},p_i)$ is the probability that a vehicle at $r_k\angle{\phi_k}$ experiences no interference from beam $b_i$.

Given a particular deployment layout $\mathcal{B_L}$ (see also Table~\ref{tbl:notation} for its definition), assuming that all beams are always active, the area of beam $b_k$ that is interfered by other beams can be computed by
\begin{equation}
A_I(k) = A_B - \int_0^{R_B} \int_{\Theta_1}^{\Theta_2} r_k\prod_{\substack{{i=1} %i=0 (Yes, i=1)
\\i\neq k}}^{B_N} \mathcal{P}_I(r_{i|k},\phi_{i|k},1) d\phi_k dr_k
\end{equation}
%
%{\color{red}\sout{HL: defintion of $\mathcal{B_L}, A_B, B_N$?}}
where $r_{i|k}\angle{\phi_{i|k}}$ is the location relative to $b_i$ given that it is at $r_k\angle{\phi_k}$ relative to beam $b_k$.

For an interference-free beam serving a vehicle, the provided service time period depends on the vehicle displacement in the beam. However, when a beam is interfered by other beams, its interference-free region is reduced, and the vehicle may experience a reduced service period due to interference. The service period is reduced to the period that the vehicle is not experiencing any interference while travelling across the serving beam.

To derive the reduced service time period, we first determine the trajectory of the vehicle within beam $b_k$ characterized by its starting point $r_k\angle{\phi_k}$ and moving direction $\psi_k$. Since the moving direction is fixed, trajectory is a straight line. By adding the starting vector $r_k\angle{\phi_k}$ and the movement vector $d\angle{\psi_k}$, the trajectory of the vehicle mobility can be computed using the following parametric equations
\begin{equation}\label{eq:trajectory}
\begin{array}{ll}
r_k(d) = \sqrt{r_k^2 + d^2 + 2r_k d \cos(\gamma_k)} \\
\phi_k(d) = \phi_k + \textrm{atan2}(d\sin(\gamma_k),r_k+d\cos(\gamma_k))\\
\end{array}
\end{equation}
where $\gamma_k=\psi_k-(\theta_k+\phi_k)$, and $r_k(0)\angle{\phi_k(0)}$ is the starting location relative to beam $b_k$, $r_k(d)\angle{\phi_k(d)}$ is the location after travelling $d$ distance. Likewise, at any location $r_k(d)\angle{\phi_k(d)}$, we denote $r_{i|k}(d)\angle{\phi_{i|k}(d)}$ to be the translated location relative to beam $b_i$.

Given a trajectory within beam $b_k$, say $\zeta$ characterized by $(r_k,\phi_k,\psi_k,b_k)$, we can determine the length of the trajectory $\tau_\zeta$. The derivation of $\tau_\zeta$ is already given in Appendix \ref{appendix:Jk} for various cases. Then the distance that the vehicle travelled experiencing no interference is given by
\begin{equation}
d_{\zeta}^{(r_k,\phi_k,\psi_k,b_k)} = \int_{0}^{\tau_\zeta} \prod_{\substack{{i=1}%i=0 (yes, i=1)
\\i\neq k}}^{B_N} \mathcal{P}_I\left(r_{i|k}(d),\phi_{i|k}(d),p_i\right) dd.
\end{equation}

Recall that we define $J_k(r_k,\phi_k,l)$ to be the probability that the \emph{service distance} of beam $b_k$ for a vehicle starting at $r_k\angle{\phi_k}$ is shorter than $l$. Considering the highway scenario with interference from other beams, we revise $J_k(\cdot)$ and denote it as $\breve{J}_k^*(\cdot)$.

Given $d_{\zeta}^{(r_k,\phi_k,\psi_k,b_k)}$ which is the distance travelled by a vehicle without experiencing interference, we can derive $\breve{J}_k^*(r_k,\phi_k,l,\psi_k)$ directly by
\begin{equation}
\label{eq:J4}
\breve{J}_k^*(r_k,\phi_k,l,\psi_k) = \left\{
\begin{array}{ll}
      1 & d_{\zeta}^{(r_k,\phi_k,\psi_k,b_k)}\leq l \\
      0 & \text{otherwise.} \\
\end{array} 
\right.
\end{equation}

Rewriting \eqref{eqn:Fstar} using $\breve{J}_k^*(\cdot)$, we yield
\begin{equation}\label{eqn:BreveFstar}
\breve{F}_L^*(l) = \frac{\int_{\Theta_{1}}^{\Theta_{2}} \int_{R_n}^{R_f} r_k \breve{J}_k^*(r_k,\phi_k,l,\psi_k) dr_k d\phi_k}{\int_{\Theta_{1}}^{\Theta_{2}} \int_{R_n}^{R_f} r_k dr_k d\phi_k}
\end{equation}
which describes $F_L^*(l)$ for highway scenario capturing the interference from other beams.

%===============================

\section{Multi-Agent Contextual Bandit for Interference Management} \label{sec:C-MAB}

We recognize that managing transmission interference in a dense small cell deployment such as one we consider in this paper requires precise knowledge of the interference condition so that beam transmissions can avoid competing and interfering with each other. Designing an algorithm that allows beams to coordinate with each other precisely is challenging since each deployment scenario has a different interference condition, hand crafting an algorithm to suit each deployment scenario is not only tedious but also prone to error due to environment change. In this paper, we present an online reinforcement learning algorithm that can not only self-learn the interference condition but also adapt to the changing environment.

We extend  
contextual multi-armed bandit (C-MAB) to enable context learning and multi-agent coordination. While the design objective of C-MAB is to learn and discover the option given a specific context, our design has slightly different objective. Instead of discovering the best option, our ML agent attempts to learn and cluster the contexts into two groups in an unsupervised manner, one of which is deemed to cause interference if a transmission is made, and another is deemed to yield an interference free transmission. With the proper classification of contexts, during the exploitation, the ML agent will only initiate a transmission when the context is classified under interference free group, or perform backoff for context classified under interference. Next, we shall describe our proposed Multi-Agent Context Learning algorithm (MACOL).

%In addition, the performance of such an algorithm is investigated in terms of its regret calculated by the difference between total mean reward that would have been achieved if the best action was selected and the total reward of the actions selected by the learning algorithm during the learning phase. 
%....\footnote{\color{red}Provide brief description the purpose of the regret.-- Done }

\subsection{Proposed Algorithm}

In a classical MAB algorithm, machine learning agent explores possible options known as \emph{actions} or \emph{arms} to learn the outcome of each option. The outcomes are often quantified into measurable \emph{rewards}, and the objective of MAB is to maximize the rewards which equates to better decision making. As the environment changes, so are the outcomes and rewards. In MAB, the rewards are often assumed to follow some probability distribution function. As a result, learning the statistical characteristics of the reward distribution is important to make optimal decision in order to maximize the overall rewards. MAB also assumes that the environment remains in the same state during the entire operation. Recognizing the underlying environment may also undergo different phases or states, and reward distribution in each state of the environment may exhibit a different statistical characteristic, it is thus necessary to observe the environment state and apply different treatments for different environment states.

Contextual MAB (C-MAB) addresses this issue by introducing the concept of \emph{contexts} into the classical MAB. In C-MAB, different contexts represent different observed environment states. For observed
environment states (contexts),  C-MAB learns the reward distribution for each environment state during exploration, associates the learned reward distribution to the observed state, and applies corresponding reward distribution for decision making during exploitation.

\subsubsection{Multiple Agents and Contexts}

A necessary step in C-MAB design is to define the environment contexts. In our design, we use the operating status of all beams in the BS cluster within the deployment scenario to form the context. The operating status of a beam can either be \emph{active} or \emph{idle}. When a beam is serving a vehicle, it is said to be in an active status, otherwise it is said to be idle as it is not servicing any vehicle. Let $\vec{S}(t)$ be a context vector collecting the operating status of all beams at the observing time $t$, and define as follows.
\begin{equation}
    \vec{S}(t) = \langle s_1(t), s_2(t), \cdots, s_{B_N}(t) \rangle, 
\label{eq:stateVectors}    
\end{equation}
where $s_i(t)\in\{0,1\}$ is the 
element describing the operating status of beam $b_i$ at time $t$ with 0 indicating idle and 1 indicating active.

Our proposed MACOL uses multiple agents for coordination. Each beam in the BS cluster has an ML agent which independently observes the contexts and interacts with the environment. We call ML agent $i$ for the agent associated with beam $b_i$. Each ML agent coordinates with others through sharing of its operating %{\color{red}\sout{HL: instantaneous?}} 
status in a form of contexts. 

The objective of the learning for each ML agent is to avoid transmission interference by coordinating transmissions among the ML agents, resulting in longer vehicle service period. 
For each ML agent, it
is sufficient %HL for each ML agent 
to observe the status of those beams that will cause interference to its transmission rather than all beams. In other words, instead of applying the observed context $\vec{S}(t)$ in the learning, each ML agent masks out the status of non-interfering beams from $\vec{S}(t)$ to form its own version of contexts. The purpose of masking out non-interfering beams is to reduce the context space for each ML agent for more focused learning and improved learning convergence. Based on the underlying deployment layout, the non-interfering beams for each beam can be easily identified, and a corresponding mask setting can be applied. The applied mask will remain unchanged throughout the operation unless the underlying deployment layout has changed. 
The masking vector indicating non-interfering beams for beam $b_i$ can be defined as follows.
\begin{equation}
\vec{M}_i = {\langle}m_{1|i}, m_{2|i}, \cdots, \ m_{B_N|i}{\rangle}, 
\label{eq:nonMask}
\end{equation}
where $m_{k|i} \in \{1,0\}$ indicates whether activation of beam $k$ does not interfere beam $i$, i.e., $m_{k|i}$ becomes $1$ when beam $k$ can interfere beam $i$ and $0$ for otherwise.
We denote $\vec{S}_{i}(t)$ %$\vec{S}_i(t)$ 
to be the observed context of beam $b_i$ after masking all non-interfering beam status to 0. 
By using (\ref{eq:nonMask}), $\vec{S}_{i}(t)$ can be expressed as follows. 
\begin{equation}
%\vec{S}_{i}(t) = &  \{ \vec{S}(t) \otimes  M_i \}  \setminus  s_i(t),  \\
%= & \{ {s}_1(t), \cdots, {s}_{i-1}(t), {s}_{i+1}(t), \cdots, {s}_{B_N}(t)  \}, 
\vec{S}_{i}(t) = \vec{S}(t) \circ \vec{M}_i %, 
= \langle s_{1|i}, \cdots, s_{k|i}, \cdots,s_{B_N|i}\rangle 
\label{eq:contextVec}
\end{equation}
where $\circ$ is the Hadamard Product operator and ${s}_{k|i}(t)$ is equal to $1$ only if beam $k$ (one of interfering beams to beam $i$) is activated. For ${s}_{k|i}(t) =0 $, it can indicate two cases, (i)~beam $k$ is not activated (while beam $k$ can interfere to beam $i$) or (ii)~the status of beam $k$ does not affect to beam $i$. This
context vector is an important information enabling the coordination among ML agents to avoid transmission interference.

\subsubsection{Actions, Rewards and Context Learning}

In classical bandit algorithms, the ML agent picks an arm using a particular policy with a model, observes the reward from the environment, and refines the model based on the observed reward. The aim of the ML agent is to learn which arm can produce the best reward. If contexts are used, the arm picking decision will also be based on the observed context. 
To apply contextual MAB to our problem, we need to design appropriate actions and rewards. We first discuss the actions. When a beam is available for the next transmission service, the ML agent can greedily choose to serve a vehicle. However, this greedy option might cause its transmission to interfere with other ongoing transmissions, thus it might choose not to perform a service and backoff for some period of time. As a result, the ML agent has two arms, one of which is to serve a vehicle and another is to perform a backoff.
% \textcolor{red}{, which can be represented as follows.}\footnote{\textcolor{red}{Can we add this eq? I am a bit hesitant, but try to add for lazy reviewers. Plz remove it if you don't like.}\color{blue}I thought this is too straightforward and not good use of space} \textcolor{red}{
% \begin{equation} \nonumber
% \mathcal{A} =\begin{cases}
% \text{serving a vehicle} & \text{if }  $a=1$\\
% \text{performing a backoff} , & \text{if } $a=0$.
% \end{cases}
% \end{equation}
% }

We shall now discuss the rewards. The ML agent learns the goodness of its action by observing the reward. In our problem, the reward should indicate whether a transmission has suffered an interference. This can be done by measuring the quality of the transmission during the transmission service. There are several ways we can measure the quality of a transmission, for example by measuring Reference Signal Received Power (RSRP), Reference Signal Received Quality (RSRQ), Received Signal Strength Indicator (RSSI), or the goodput rate in bps measuring the average number of bits per second successfully received by the vehicle over the transmission period. In this work, we shall use the goodput rate for the reward. A higher goodput rate indicates good channel quality whereas a lower goodput rate may be due to poor channel quality or transmission interference, or both.

Given the above design of actions and rewards defined for our problem, we see that a classical MAB is unsuitable to apply to the problem. In a classical MAB, a reward is observed after take an action. If no action is taken, no reward can be observed. In our problem, as the ML agent can choose to backoff, no transmission will take place and thus it is impossible to obtain a reward if the backoff action is chosen. It is also not trivial to assign a reward to a backoff action as this requires other ML agents to confirm that a backoff action has indeed avoided an interference and quantify the benefit. To overcome these issues, we introduce a key process into a classical MAB. We add an unsupervised learning model into MAB for the ML agent to learn whether a transmission will cause interference with others. Precisely, based on the contexts received from the environment, the ML agent learns  which contexts will cause interference to others and which will not. In other words, the ML agent classifies contexts into two groups, one labeled as ``transmission-interference" group and another as ``interference-free" group.

The classification of the contexts relies on the observed rewards after each transmission is made. Upon completion of a transmission service, a reward is produced. This reward is associated with the context at the time when the decision of performing a transmission service is made. We further define the value of each context to be the average reward associated with the context. As each context has a value, we can classify each context to be either ``transmission-interference" or ``interference-free" based on its value. With one-dimensional data, we can simply rank contexts based on their value, set a threshold to separate them into two groups. Those in the higher (resp. lower) value group signify ``interference-free" (resp. ``transmission-interference"), as a higher reward indicates a non-interfering transmission, and a lower reward indicates a transmission suffering interference or poor channel condition.

With our design, we not only avoid the need to quantify a reward for a backoff action, but also achieve the coordination among all ML agents by using contexts to share status. Unlike classical contextual bandit algorithms that learn directly which action is best given a context, our algorithm learns to classify contexts containing inputs from other ML agents, and take appropriate action based on the classified outcome. Hence we call our bandit algorithm the Multi-Agent Context Learning (MACOL) algorithm.

Let $\mathcal{C}_i(t)$ denote a set containing all contexts already observed by beam $b_i$ at time $t$, and these contexts carry a valid reward value. When a transmission service is initiated at beam $b_i$, context, $\vec{S}_i(t)$, is immediately observed. Beam $b_i$ continues to serve the vehicle for $\Delta t$ period until it has left the coverage of beam $b_i$, then the reward is measured and the average reward value of context $\vec{S}_i(t)$ is updated. Let $\rho_i(t+\Delta t)$ be the measured reward which is associated with context $\vec{S}_i(t)$, the average reward value of context $\vec{S}_i(t)$ is updated by 
\begin{equation}
\bar{\rho}_{\vec{S}_{i}}(t+\Delta t) = \frac{\bar{\rho}_{\vec{S}_{i}}(t) \cdot k_{\vec{S}_{i}}(t) + \rho_i(t+\Delta t)}{k_{\vec{S}_{i}}(t)+1}, 
\label{eq:increAvg}
\end{equation}
where $k_{\vec{S}_{i}}(t)$ is the trial number of transmission actions under context $\vec{S}_{i}(t)$. Besides, the set of observed contexts of beam $b_i$ is also updated accordingly by $\mathcal{C}_i(t+\Delta t) = \mathcal{C}_i(t) \cup \{\vec{S}_i(t)\}$.

After providing a service, beam $b_i$ is ready for the next service. It first observes the context and performs classification of the observed context. It proceeds with a transmission service if the observed context is classified as ``interference-free", otherwise, a backoff is performed. As mentioned earlier, the ML agent ranks all already seen contexts based on their reward value, and set a threshold to partition those above the threshold to be ``interference-free". In our design, we use the average context value as the threshold. That is,
\begin{equation}
\dot{\rho}(t+\Delta t) = \frac{1}{|\mathcal{C}_i(t+\Delta t)|} \sum_{{\vec{S}_{i}(t+\Delta t)}\in \mathcal{C}_i(t+\Delta t)}{\bar{\rho}_{\vec{S}_{i}(t+\Delta t)}}, 
\label{eq:avgReward}
\end{equation}
where $\dot{\rho}(t+\Delta t)$ is the average context value over all observed contexts.

In the case that a backoff action is performed, the ML agent remains silence for some period before resuming back to the normal operation. The backoff duration can be as simple as a predefined constant value. Here we use an adaptive duration for the backoff. The backoff duration is set to the average connection duration observed for the current context.

\subsubsection{Exploration and Exploitation}

Algorithm~\ref{alg:cmab} describes our proposed MACOL algorithm for ML agent $i$. 
Our algorithm uses the explore-first strategy including two phases, firstly exploration and then exploitation.
During the exploration phase of the predefined period, we let the ML agent explore and learn the environment for a specific duration. 
To do this,
ML agent $i$ continuously serves vehicles within its coverage area one at a time. Random vehicle selection is used when there are more than two vehicles to choose for a service. The ML agent only manages the beam of a particular frequency band. The entire frequency band is allocated to the vehicle, and thus the ML agent only manages one vehicle at a time. 

The service is said to be completed whenever a connected vehicle has departed the beam. Then the ML agent computes and updates the reward for the context accordingly using (\ref{eq:increAvg}) (lines 30-34 in Algorithm~\ref{alg:cmab}). 

As the exploration period expires, ML agent $i$ immediately switches to exploitation mode. During exploitation, for an observed context $\vec{S}_{i}(t)$ (derived from $\vec{S}(t) \circ \vec{M}_i$) at time $t$,
the ML agent updates the classification threshold and checks the belonging of the context (lines 16-17). If the average reward value of the context is higher than the threshold the context is classified as ``interference-free", and the ML agent proceeds to perform the transmission service action (lines 18-21). Otherwise, it performs the backoff action (line 23). Note that the ML agent may occasionally perform exploration with $\epsilon$ probability (lines 15, 26).

\begin{algorithm}
\caption{Context Learning MAB for Interference Management for ML agent $i$}\label{alg:cmab}
\hspace*{\algorithmicindent} \textbf{Input:}\\
\hspace*{\algorithmicindent}\hspace*{\algorithmicindent}The context vector $\vec{S}_i$ after masking.\\
\hspace*{\algorithmicindent} \textbf{Initialization:}\\
\hspace*{\algorithmicindent}\hspace*{\algorithmicindent}$\mathcal{C}_i \gets \emptyset$.\\
\begin{algorithmic}[1]
\Procedure{Exploration}{}
\If{No vehicle is available in the beam}
    \State return
\EndIf
%\If{ \textcolor{red}{$rand()$ is larger than $\epsilon$}} 
\State Select a vehicle randomly
\State Serve the vehicle until it moves out of the beam
\State Compute the reward $\rho_i$
\State UpdateReward($\vec{S}_i,\rho_i$)
\EndProcedure
\\
\Procedure{Exploitation}{}
\If{No vehicle is available in the beam}
    \State return
\EndIf
\If{ rand() is larger than $\epsilon$}
\State $\dot{\rho} = \frac{1}{|\mathcal{C}_i|} \sum_{s\in \mathcal{C}_i}{\bar{\rho}_{s}}$
\If{$\bar{\rho}_{\vec{S}_i}>\dot{\rho}$}
\State Select a vehicle randomly
\State Serve the vehicle until it moves out of the beam
\State Compute the reward $\rho_i$
\State UpdateReward($\vec{S}_i,\rho_i$)
\Else{}
\State Wait until the backoff process ends
\EndIf
\Else{}
    \State Exploration()
\EndIf
\EndProcedure
\\
\Procedure{UpdateReward}{context $\vec{s}$, reward $\rho$}
\State $\bar{\rho}_{\vec{s}} \gets \frac{\bar{\rho}_{\vec{s}} \cdot k_{\vec{s}}+\rho}{k_{\vec{s}}+1}$ \Comment{$\bar{\rho}_{\vec{s}}$ is set to 0 initially}
\State $k_{\vec{s}} \gets k_{\vec{s}}+1$ \Comment{$k_{\vec{s}}$ is set to 0 initially}
%\If{$\mathcal{C}_i \cap \{\vec{s}\}=\emptyset$}
\State $\mathcal{C}_i \gets \mathcal{C}_i \cup \{{\vec{s}}\}$
%\EndIf
\EndProcedure
\end{algorithmic}
\label{MABalgo}
\end{algorithm}

%vvvvvvvvvvvvvvvvvvvv

\section{Simulation and Result Discussion}
\label{sec:result}

\subsection{Scenario Setup}

In this section, we present the simulation results to show the performance of our proposed MACOL algorithm. We use our own developed Python Mobility Simulation Platform (PyMoSim)
%\footnote{We plan to release the source code of PyMoSim Platform in the near future.}
for the simulation. Our main purpose is to confirm that the MACOL ML agent can indeed learn whether initiating a transmission by activating its beam will potentially cause an interference to other transmissions by neighbouring beams, 
and use this learned knowledge to avoid interference.

\begin{figure}[!pt]
    \centering
    \includegraphics[width=0.95\columnwidth]{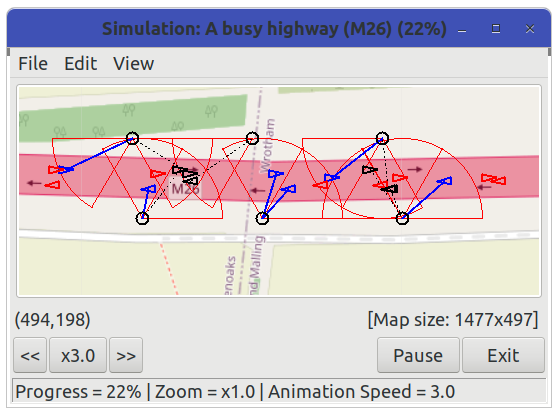}
    \caption{Screen snapshot of Pymosim simulation running our proposed highway scenario {based on geometric framework}. Simulation code is given in https://github.com/cfoh/beam-analysis}
    \label{fig:sim-screenshot}
\end{figure}

We focus on a highway scenario illustrated in Fig.~\ref{fig:roadlayout} including 6 BSs having 3 mmWave beams. In Fig.~\ref{fig:sim-screenshot}, we show a screen snapshot of our simulation.
In our study, we use ideal sector-shaped radiation pattern with the radiation range of 80 m and beamwidth of $60^{\circ}$. The ideal sector-shaped radiation pattern is illustrated in Fig.~\ref{fig:sim-screenshot}.
While a BS consists of 3 beams, these 3 beams are independently controlled and operated. They are mainly used for macro BS to offloading infotainment data to vehicles \cite{Kose2021}. Whenever the beams are available, they may decide to serve a passing vehicle. %depending on the set strategy. 
When they decide to serve a passing vehicle, we assume that the downlink infotainment data 
can be quickly channelled to the BS to transmit the data to the served vehicle.
During the downlink transmission, the amount of data successfully transmitted will be observed to calculate the goodput rate as the reward. Assuming full downlink buffer rerouted from the macro BS to the local small cell BS, the downlink transmission is performed until the vehicle has departed the beam (i.e., the connection is lost). For transmission to service vehicles, 
the beam will allocate a specific frequency band. % to transmit the data to the vehicle until the connection is lost.
The BS may serve multiple vehicles using multiple beams, each with an orthogonal frequency band, our experiment focuses on a specific frequency band used by all vehicles and BSs.
That is, neighbouring beams can interfere each other and only one vehicle can be served by a beam at one time.

We consider different traffic load conditions for the highway scenario, from the light traffic load of 6 vehicles to the heavy traffic load of 30 vehicles. To ensure we maintain the constant number of vehicles on the highway for a specific experimental setup, in our simulation, we create equal amount of vehicles from the two edges of the highway section, and set a random speed for each vehicle to travel across the highway section. Once the vehicle has reached the other end of the highway section, it immediately reappears at its original start location as a new vehicle and start to travel immediately with a new speed randomly set. The speed is set to between 80 km/h and 110 km/h which is the typical travelling speed on highway in most countries.

When a vehicle travels across the highway section, it may receive a service from a beam without any interference, it may be served but experiencing interference from other nearby beams, or it may not be served due to all nearby beams being busy. 
To evaluate the impact of such interference on transmission to vehicles, 
we are particularly interested in the ratio of overall service period to the entire period that a vehicle experienced while passing the segment of the highway. That is, if a vehicle takes $T_E$ period of time to pass the entire segment of the highway, and during its journey, it receives transmission service for $T_S$ among of time, then the ratio of service period to the entire period is $\frac{T_S}{T_E}$. Additionally, we also measure the ratio of its interference duration and outage duration to the entire period.

Three beam allocation algorithms are considered in our testing. The first method is the random vehicle selection. Whenever a beam is available to provide service, and vehicles are found within its beam coverage, the beam randomly selects a vehicle to provide service. In the second method, rather than randomly select a vehicle, the BS measures the uplink transmission signal strength and allocates the service to the vehicle with the strongest signal. This is also known as Best SNR which is a widely used strategy. With two reference schemes, our proposed MACOL algorithm is evaluated. 

{%\color{blue}
In the following experiments, we first study the impact of interference on service distance. We use geometric framework in this study to show the numerical and simulation results. We next investigate the effectiveness of MACOL for interference management where both geometric framework and practical model are considered. Finally, we study the learning efficiency and computational complexity of our proposed MACOL.
}

\begin{table}[!htbp]
\caption{System parameters setting}
\begin{tabular}{  m{3.4cm} | m{4.6cm}  }
 \hline
Parameter & Value  \\ \hline \hline
Target scenario  & DL transmission in a highway covered by mmWave small cell BSs \\ \hline
no. of BSs  & $6$ \\ \hline
no. of mmWave beams per BS & $3$ \\ \hline
beam transmission range and beamwidth  & $80$ m \& $60^o$ \\ \hline
%max no. of neighbouring beams (of a specific beam) & 17 \\ \hline
max no. of interfering beams & 5 \\ \hline
%channel model  & xxxx \\ \hline
no. of vehicles & varying in [$6,40$] \\ \hline
vehicle traveling speed  & random in [$80, 110$] km/h when entering the highway segment, remains unchanged while passing the highway\\ \hline
overall simulation time  & 2000 sec \\ \hline
exploration period  & one of 120, 180, 240, 300, 600 sec  \\ \hline
$\epsilon$ (in the exploitation phase) & $0.05$ \\ \hline
no. of observable contexts & 32 \\ \hline
\end{tabular}
\label{tab:sysParams}
\end{table}

\subsection{Impact of Interference on Service Distance}

\begin{figure}[!pt]
    \centering
    \includegraphics[width=0.86\columnwidth]{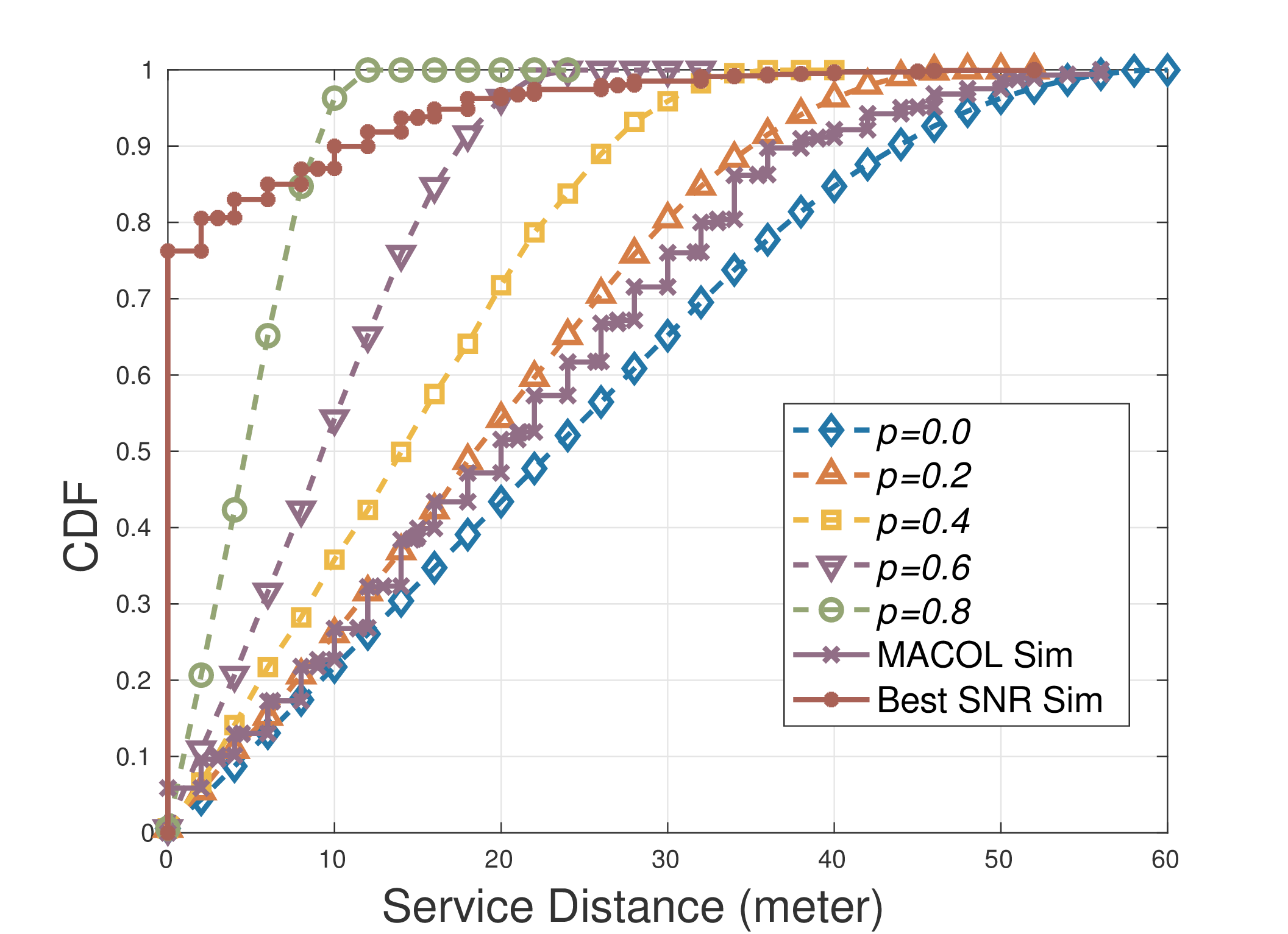}
    \caption{Illustration of numerical service distance CDF for various $p$ settings and simulated service distance CDF for MACOL and BestSNR techniques with traffic load of 30 vehicles,
    where $p$ is the probability that its neighbouring beam is active during a transmission service by a beam. }
    \label{fig:cdf-service-distance}
\end{figure}

{%\color{blue}
Considering the interference for geometric framework, }
in \eqref{eqn:Fstar}, we derive the probability that a vehicle travelled a distance of $l$ or less within a beam in the highway scenario. By applying \eqref{eq:J4} to \eqref{eqn:Fstar}, we can focus only on the service distance which is the distance including only when interference-free transmission service is received by the vehicle. Fig.~\ref{fig:cdf-service-distance} plots the numerical results of service distance CDF for various $p_i$ settings where $p_i$ is the probability that neighbouring beam $b_i$ is active during an observation. In our numerical computation, we set all $p_i$ to carry the same value of $p$. Note that the derivation of \eqref{eqn:Fstar} assumes a saturated traffic load condition where when a service is completed and the beam has decided to initiate a new service, the highway is full of vehicles in all locations within the beam coverage, and the beam will unbiasedly select one vehicle for service which can be at any location. To match this high load condition in our simulation, we use traffic load setting of 30 vehicles. In Fig.~\ref{fig:cdf-service-distance}, we plot the simulated service distance CDF for MACOL and the Best SNR techniques.

We first focus on the case of $p=0.0$ which is equivalent to the case of no interference. The CDF result represents the best service distance performance which indicates the upper-bound service distance performance. We see that our simulated service distance CDF for MACOL resides between cases of $p=0.0$ and $p=0.2$ showing that our proposed MACOL can indeed maintain low interference equivalent to around $0.1$ of probability of interference. As $p$ increases, the service distance for the same percentile reduces. We also include the simulated  
service distance CDF for the Best SNR technique for comparison. We see that over $75\%$ of the connections experience total interference with zero service distance, over $85\%$ fall below that of $p=0.8$, and over $95\%$ fall below that of $p=0.6$. The results suggest very intensive interference due to its greedy transmission characteristics. %{\color{red}\sout{HL: might be useful to include no. of vehicles indicating traffic loads for Fig.3, can we change y-axis label to "CDF" from "Probability"?}\footnote{\color{blue}traffic load added to the caption. CDF is the curve, not the y-axis, although many papers use CDF for y-axis, it's more accurate to use prob.}}

\begin{figure}[!pt]
    \centering
    \includegraphics[width=0.86\columnwidth]{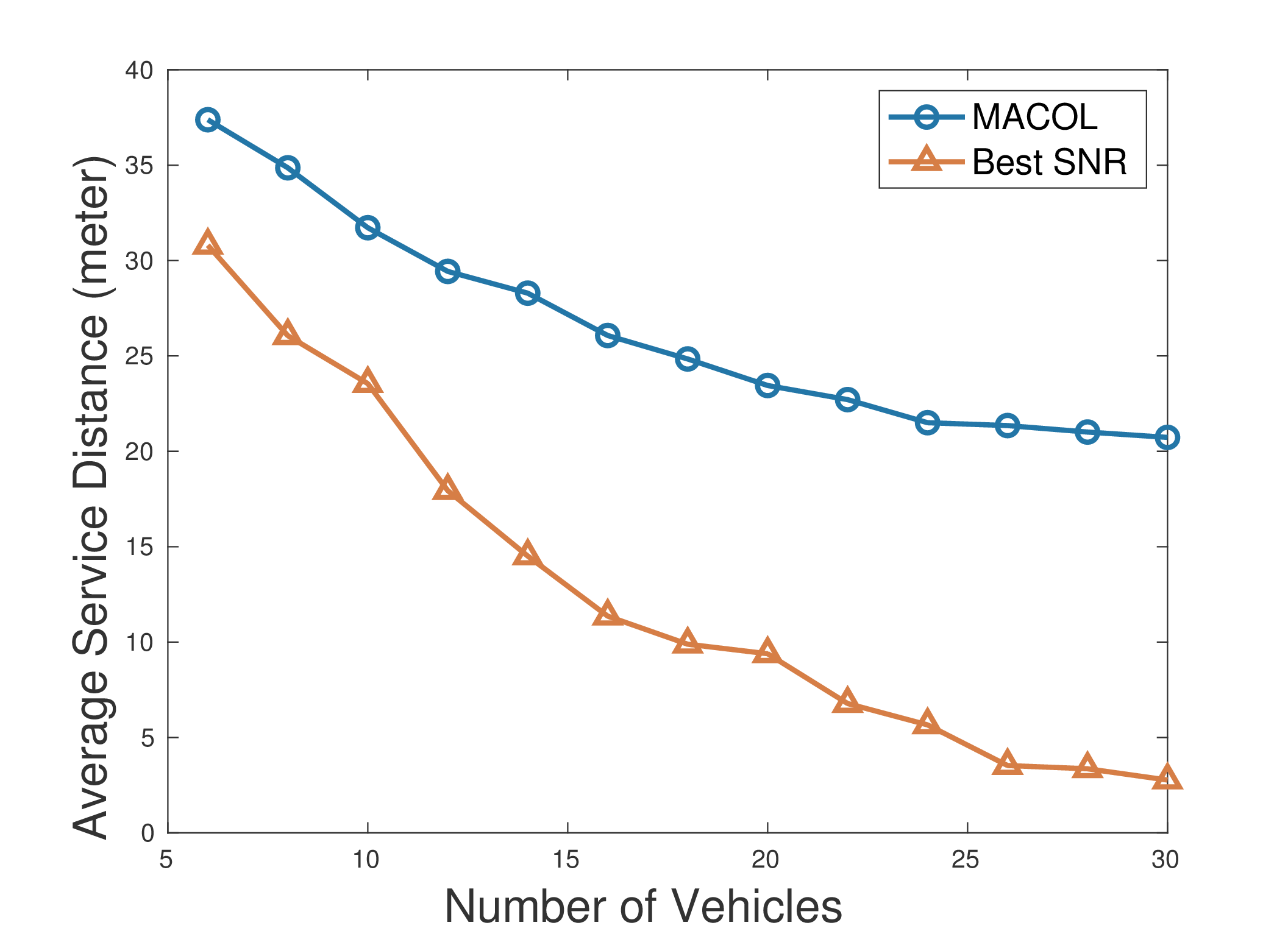}
    \caption{Mean service distance for MACOL and Best SNR techniques with various traffic load conditions.}
    \label{fig:sim-service-distance}
\end{figure}

Next, we compare the mean service distance for our proposed MACOL and Best SNR techniques over a range of numbers of vehicles in Fig.~\ref{fig:sim-service-distance}. From the figure, we see that with light traffic load of 6 vehicles, both algorithms yields over 30m of mean service distance. As the traffic load increases, we see that the Best SNR algorithm drops, and the gap between the two algorithms continue to widen. On the other hand, for the mean service distance of MACOL, the drop is relatively slow. Interestingly, it stabilizes at above 20m when traffic load becomes intense suggesting effective management of interference even under very high traffic load conditions.

\subsection{Effectiveness of MACOL for Interference Management}

\begin{figure}[hbt]
    \centering
    \includegraphics[width=0.86\columnwidth]{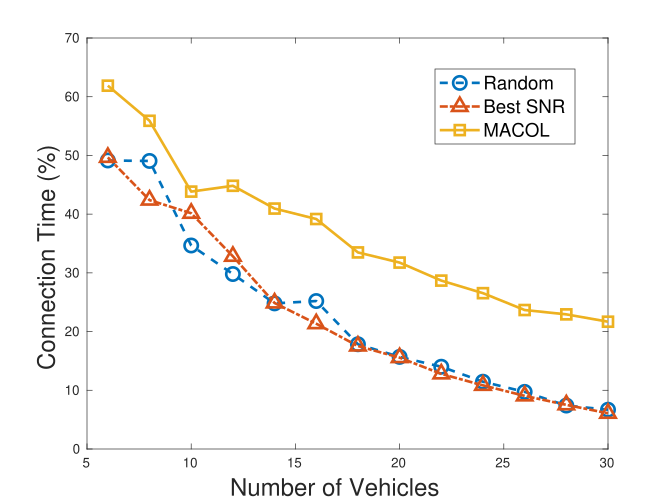}
    \caption{Percentage of service period experienced by a vehicle passing the highway.}
    \label{fig:service-time}
\end{figure}

\begin{figure}[hbt]
    \centering
    \includegraphics[width=0.86\columnwidth]{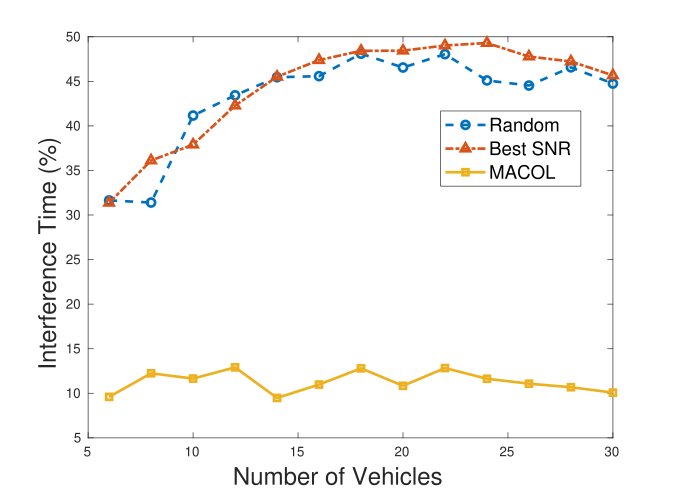}
    \caption{Percentage of interference time experienced by a vehicle passing the highway.}
    \label{fig:interference-time}
\end{figure}

We shall now focus on the effective of our proposed MACOL interference management. {%\color{blue}
We first consider the interference for geometric framework.} We vary the traffic intensity conditions from low to high and compare the service period ratios of a passing vehicle with various methods in Fig.~\ref{fig:service-time}. As can be seen, the service period ratio of a vehicle drops as the traffic condition becomes more intensive. A vehicle receives communication services over 50\% of the time while traveling across the highway when there are 6 vehicles on section of the highway. As the number of vehicles increases, the communication services for each vehicle drops. There are two reasons contributing to the drop. Firstly, as the communication services are shared among all vehicles, with more vehicles on the highway, each vehicle receives reduced service duration. Secondly, more vehicles cause increased communication interference leading to lower service duration. While the drop in service period ratio is inevitable due to the sharing of the same frequency band, our proposed MACOL manages to mitigate the interference even with increasing traffic intensity. We see in Fig.~\ref{fig:interference-time} that MACOL maintains interference at around 10\% even when the traffic intensity increases, whereas growing interference is seen for other methods when traffic intensity increases. The ability to maintain interference when traffic intensity increases shows that our proposed MACOL is effective in managing the interference.

\begin{figure}[hbt]
    \centering
    \includegraphics[width=0.86\columnwidth]{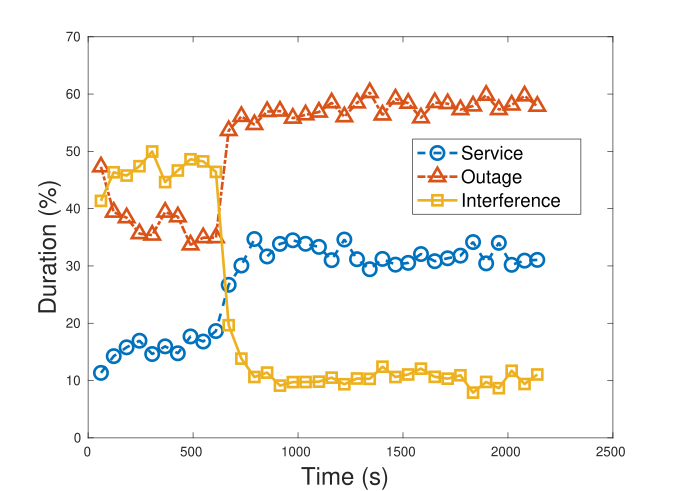}
    \caption{Service, outage and interference duration over time for 20 vehicle setup.}
    \label{fig:20-cars}
\end{figure}

To understand how MACOL manages the interference, we show the fraction of time that a vehicle receives connection service, experiences interference or outage while passing the highway in Fig.~\ref{fig:20-cars}. Here we set the number of vehicles to 20. Note that our proposed MACOL algorithm uses explore-first strategy, during the first 10 minutes (or 600 seconds), the ML agent acts greedily to learn the contexts. During this phase, we see that a vehicle typically experiences around 45\% of interference, 15\% of service period without interference, and the remaining time not being able to receive a service as all nearby beams are occupied by other vehicles. Due to the high number of vehicles, and the aggressiveness of each beam attempting to serve passing vehicles whenever possible, many pairs of vehicles are served within the same region causing high interference experienced by each vehicle. Our proposed MACOL ML agent overcomes the interference by using backoff. When the ML agent judges that a service may cause an interference, it performs backoff instead to avoid the potential interference. As a result, during the exploitation, the ML agent uses the classified contexts to identify and avoid potential interference. We see that the outage ratio increases because ML agent backing off where fewer transmissions are performed. The backing off of services leads to translation of interference to service where interference ratio drops and service ratio increases. This is clearly shown in Fig.~\ref{fig:20-cars} that when interference ratio drops by 35\% after the exploitation, outage and service ratios each jumps by over 15\%.

\begin{figure}[!pt]
    \centering
    \includegraphics[width=0.86\columnwidth]{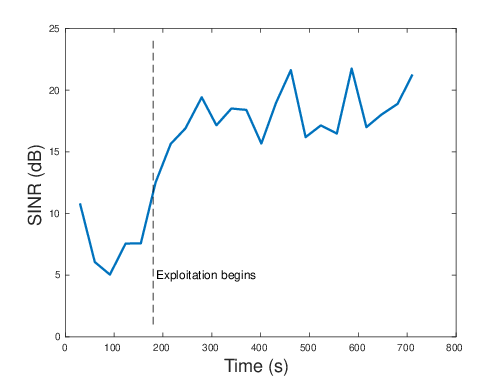}
    \caption{{Transmission quality measured in SINR of a beam service over time for 20 vehicles setup.}}
    \label{fig:sinr}
\end{figure}

{%\color{blue}
We now consider the interference model for practical setup. In this setup, the channel model described in \eqref{eqn:pathloss} is used instead, and SINR is used to determine a decodeability of a transmission. In this case, instead of simply classifying a transmission to either an transmission-interference or interference-free, we measure the transmission quality using SINR. In Fig.~\ref{fig:sinr}, we plot the SINR of transmissions from a specific beam over the time. The beam we pick is the the north pointing beam located at the south of the highway. As can be seen, the transmission SINR is below 10dB during exploration. After the learning, the transmission SINR jumps to around above 15dB. This again confirms the effectiveness of the interference management of MACOL.
}

\subsection{Transmission Reliability}

\begin{figure}[!pt]
    \centering
    \includegraphics[width=1.0\columnwidth]{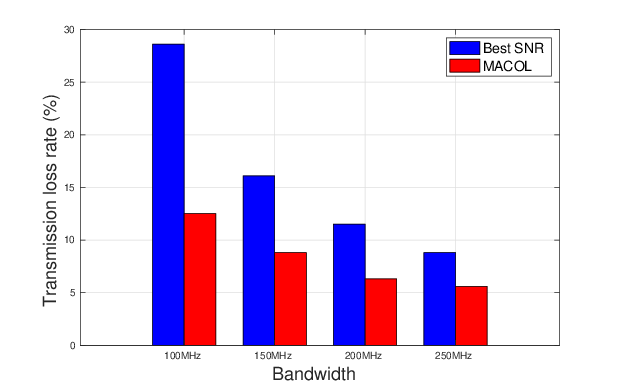}
    \caption{Impact of bandwidth on transmission loss rate for Best SNR and MACOL for 10 vehicles setup.}
    \label{fig:txloss}
\end{figure}

{
%\color{blue}
To support the effective V2X communication, one of the requirement would be the transmission reliability. Depending on the ITS use case and application, the required levels of transmission reliability can be varied. For example, Cooperative Awareness applications like forward collision warnings typically demand reliability levels in the range of 90-95\%, whereas Cooperative Sensing applications such as crash mitigation require an even higher reliability threshold of 99\%~\cite{boban2017use}.
%For example, 3GPP considers 90\% reliability (packet error rate) to support an ITS use case, which is the update trajectory plan for the lane change manoeuvre \cite{TR22.806}. 90\% reliability is to ensure low interference such that over 90\% of the transmissions can be received successfully.
In this experiment, we study the performance of MACOL for transmission reliability measured by the loss rate due to interference under different bandwidths. As given in Table~\ref{tab:txParams}, the mmWave BS uses Frequency Range 2 (FR2) 50MHz bandwidth for transmission to serve a vehicle. For a given number of vehicles travelling on the highway, by increasing the bandwidth to $n\times 50$MHz, the BS can serve $n$ vehicles simultaneously on different frequency bands, and this offers a means to reduce interference. As $n$ increases, transmission interference reduces and reliability improved. %Our focus here is to study the appropriate $n$ that can support at least over 90\% transmission reliability to support the requirement of some of the ITS use cases.
We have implemented this support in our simulation and measured the transmission loss rate for a given transmission. The results are presented in Fig.~\ref{fig:txloss} for Best SNR and our proposed MACOL solution. As can be seen,  Best SNR has over two times higher transmission loss rate than MACOL at 100MHz bandwidth. To support over 90\% of transmission reliability, while Best SNR requires 250MHz (or 5 blocks of 50MHz frequency bands) to operate, our proposed MACOL algorithm only requires 150MHz (or 3 blocks of 50MHz frequency bands) to match the performance, representing a bandwidth saving of 100MHz.
%\footnote{{\color{blue}HL: While 50MHz is for a single user and $n \times 50 MHz$ will be the BW for $n$ user, for me, it is not clear the meaning of resource saving. The given text can be interpreted that MACOL can support 3 users with 150MHz while Best SNR supports 5 users with 250MHz. Would it be possible to add some more explanations? In case two of you are fine, plz ignore my comments} }.
This outcome underlines the efficacy of MACOL in enhancing reliability, particularly in scenarios where heavy interference and limited resources are consideration. However, it is crucial to acknowledge that as heavy interference setup scenario is considered in this paper, under some relaxed setup and together with interference mitigation techniques, MACOL can achieve very high reliability (99\%) to support some mission critical ITS use cases.
}

\subsection{Learning Efficiency}

\begin{figure}[!pt]
    \centering
    \includegraphics[width=0.86\columnwidth]{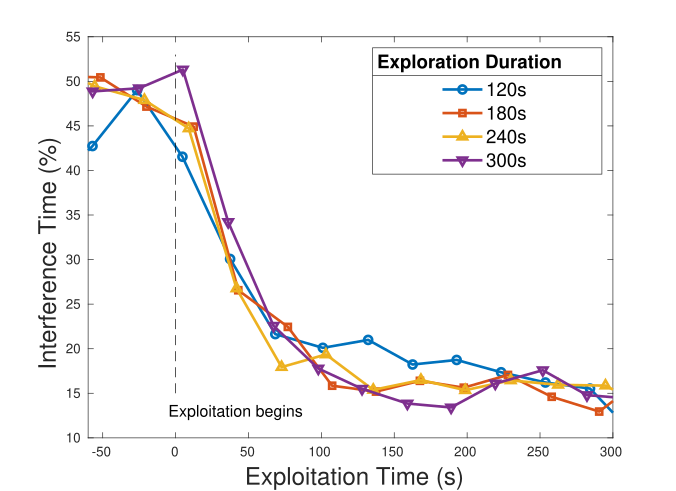}
    \caption{Comparison of learning effectiveness for different exploration duration periods. All results are aligned based on their exploitation time such that it begins at $t=0$. }
    \label{fig:exploration}
\end{figure}

%HL Our earlier regret analysis shows that...\footnote{\color{red}Haeyoung, please provide the right statement}. 
Here we are interested in how effective the ML agent learns during the exploration. Fig.~\ref{fig:exploration} illustrates the average interference time in percentage experienced by vehicles crossing the highway after the ML agent has switched to exploitation. In this experiment, we set the number of vehicles to 20. The time at 0 on the x-axis signifies the start of the exploitation phase for various exploration period settings. During the exploration phase shown before time $t=0$, vehicles experience high interference at the level of just below 50\% as all ML agents greedily learn the environment leading to high interference. After switching to the exploitation phase, each ML agent begins to coordinate by using shared context information to make decision, and hence we expect a drastic drop in the interference level. From the figure, we see that the ML agent can lower interference level quickly after switching to exploitation. The shortest exploration setting used in our experiment is 120s. Except for the 120s exploration duration setting where it drops slower than others and remains above 20\% for 120s during the exploitation period before dropping below 20\%, the other settings show similar performance in terms where the interference percentage level quickly drops below 20\% after switching to exploitation. The results show that our proposed MACOL does not require a long training period. In a busy traffic condition, several minutes are sufficient for the ML agent to learn how to manage interference which makes the algorithm practical for deployment. 

The main knowledge that the ML agent needs to learn is the contexts and their relationship to the decision outcomes. To do so, the ML agent must explore the reward of each context by serving a vehicle during when the context is reported by the environment. The number of contexts that each ML agent can observe depends on the number of neighbouring beams that may cause interference. In our setup, a beam, say at the south site, can potentially be interfered by 3 beams from the opposite side of the highway at the north site, and 2 adjacent beams on the same side of the highway at the south site. The same applies to the beam at the north site. With 5 potential interfering beam each may report a status of either active or idle, this gives $2^5=32$ different contexts. Ideally, the ML agent needs to serve a vehicle in each of these 32 contexts to establish the full picture of rewards in order to derive an appropriate threshold to cluster which contexts may cause interference and which not. In our experiment, with 120s of exploration time and 20 vehicles constantly appearing on the highway, we observe from our simulation that each beam has already served around 50 vehicles, which has provided sufficient statistics for the ML agent to classify seen contexts. Since the number of contexts is not high, with sufficient traffic on the highway, the ML agent can quickly learn to differentiate and classify contexts into interfering and non-interfering groups.

\subsection{%\color{blue}
Computational Complexity and Signaling Overhead}
%\footnote{\color{red}AK: better to use $\mathcal{O}(.)$ notation to represent the complexity states.}

{%\color{blue}

As illustrated in Algorithm~\ref{alg:cmab}, the computation of MACOL does not involve in any loop. In other words, the computational complexity of MACOL is linear with respect to the number of agents. During exploration, the ML agent greedily explore the environment independently. The computational complexity is independent of the number of ML agents, traffic load condition and number of interfering beams. During exploitation, the ML agent requires to compute the threshold using \eqref{eq:avgReward} which depends on the number of interfering beams. The number of interfering beams depends on the surrounding of the beam rather than the total number of ML agents. In our case, the largest number of interfering beams is 5. This makes our proposed MACOL solution scalable for deployment.

As a multi-agent ML algorithm, ML agents need to coordinate with each other which introduces signaling overhead. In MACOL, agents coordinate among each other by pushing their status and pulling the global context vector. Both status information and global context vector are small in size and can be represented by several bytes of information. For the uplink signaling, each agent updates its status whenever the beam changes its status between idle and active. Since the beam usually remains in active mode serving a vehicle for seconds, the changes are infrequent, and uplink signaling overhead is negligible. Likewise, for the downlink signaling, each agent pull the global context vector when a decision is required to decide whether a service or a backoff is to be performed. This will occur only when an active beam turns idle which is also infrequent. Thus the downlink signaling overhead is also very small. Together with the small size of signaling information, the signaling overhead of MACOL is small to negligible.
}

\section{Conclusion {and Future Work}}
\label{sec:conclusion}

% needed in second column of first page if using \IEEEpubid
%\IEEEpubidadjcol

In this paper, we studied the interference of a densely deployed mmWave small cell network for vehicular communications. Focusing on a highway scenario, we derived an analytical model to compute the vehicle sojourn time within a beam under the influence of interference caused by neighbouring beam transmissions, and we showed the impact of interference on the service distance performance of a vehicle while travelling across a beam.

To manage the interference, we proposed context learning bandit algorithm based on C-MAB with extension to enable learning of contexts. Different from classical MAB which aimed to maximise rewards, our proposed MACOL aimed to learn and classify the contexts using the rewards, and made decision based on the observed contexts. While our proposed algorithm used multi-agent design, it only involved in infrequent exchanges of small sized environment contexts. Simulation experiment was used to test our proposed design. Our simulation results showed that our proposed algorithm managed to maintain low interference level of around 10\% even under a heavy load condition. The simulation results were compared and the low interference level of MACOL was confirmed by our numerical results. With supports from further simulation results, we illustrated the effectiveness of interference avoidance of MACOL showing reduced interference and increased service duration during the exploitation of MACOL. 
Using a practical channel model, we presented the SINR performance showing improved SINR when MACOL the exploiting the previously learned information. 
{
%\color{blue} 
Considering transmission reliability for ITS safety applications, MACOL can reach nearly 95\% reliability at 250MHz bandwidth, whereas Best SNR can slightly exceed 90\%. Note that, MACOL only required 150MHz of bandwidth to achieve same level, which is a 40\% reduction of radio resource. 
}
Finally, we discussed the learning rate and illustrated that MACOL achieved effective operation with very low exploration iteration. Particularly in our scenario, with busy road traffic, MACOL can quickly learn and operate even with a cold start since it only requires a few minutes of exploration to learn the environment.

{%\color{blue}
In the future, we shall extend MACOL to other scenarios such as a dense mmWave small cell deployment scenario to study its effectiveness in interference management and perform further refinement to the algorithm. We shall also extend our analytical framework to include channel models. In the practicability aspect, we shall investigate the technical and regulatory challenges to bring MACOL to the real-world network, as well as the security and privacy of the solution.}

\appendices

\section{Derivation of $J_k(r_k,\phi_k,l)$}\label{appendix:Jk}

Here we derive $J_k(r_k,\phi_k,l)$ which is the probability that a vehicle travelling distance within beam sector $b_k$ is shorter than $l$. Note that beam $b_k$ is characterized by its Cartesian location $(x_k,y_k)$ and pointing direction $\theta_k$. Since the starting location of the vehicle $r_k\angle{\phi_k}$ is specified relative to $b_k$, the beam location $(x_k,y_k)$ is not required in the derivation. Additionally, if $r_k\angle{\phi_k}$ is not within beam $b_k$, we simply have $J_k(r_k,\phi_k,l)=0$.

%Fig.~\ref{fig:beamgeometry} 
Fig.~\ref{fig:beamGeom} in the earlier section illustrates the geometry layout of the beam sector and the vehicle within. The beam sector has three points $P_1$, $P_3$ and $P_4$ %{\color{red}\sout{HL: $P_2 --> P_4$?}} 
where $P_1$ is the beam origin located at $(x_k,y_k)$. The vehicle within the beam located at $P2$ may travel in any direction and may depart the beam from either the left edge $\overline{P_1P_3}$, 
%{\color{red}\sout{HL: $\overline{P_1P_3}$?}} 
the right edge $\overline{P_1P_4}$,  
%{\color{red}\sout{HL: $\overline{P_1P_4}$?}}, 
or the arc $\widehat{P_3P_4}$.  
%{\color{red}\sout{HL: $ \widehat{P_3P_4}$?}}. 
Depending on which side the vehicle departing the beam, the derivation of $J_k(r_k,\phi_k,l)$ requires different treatments. For each case, we shall derive the probability $\mathcal{P}_i(l)$ that the vehicle travelled distance is shorter than $l$ out of all possible cases. Then $J_k(r_k,\phi_k,l)$ is simply the sum of all probabilities, that is
\begin{equation}
J_k(r_k,\phi_k,l) = \sum_{i=1}^3 \mathcal{P}_i(l).
\end{equation}

In the following, we shall focus on each individual case and derive the corresponding $\mathcal{P}_i(l)$.

\begin{figure}[hbt]
\includegraphics[width=.5\textwidth,height=!]{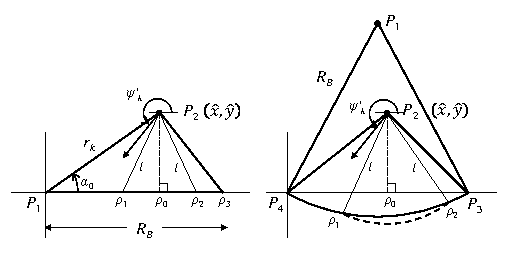}
\caption{Illustration of the transformed geometry layout for Cases 1 and 2 (left), and for Case 3 (right). For Case 1, rotation and reflection are performed in the transformation. For Case 2, only rotation is needed. For Case 3, translation and rotation are performed to bring $P_4$ to the origin and $P_3$ onto the x-axis.}
\label{fig:beamgeometry}
\end{figure}

\subsection*{Case 1: Departing From Left Edge}

We first derive Case 1 where the vehicle departs from left edge. We begin by performing geometry transformation including translation, rotation and reflection such that $P_1$ is at origin, and the edge $\overline{P_1P_3}$ of left triangle $\Delta P_1P_2P_3$ coincide on x-axis as shown in Fig.~\ref{fig:beamgeometry}. In the transformed system, point $\rho_3$ is $P_3$. We derive the location of $P_2$ as $(\hat{x},\hat{y})$ and the transformed travelling angle $\psi'_k$ %{\color{red}\sout{HL:$\psi'_k$ in Fig.9}} 
as
\begin{subequations}
\begin{align}
& \alpha_0 = \frac{\Omega_B}{2} - \phi_k \\
&\hat{x} = r_k \cos(\alpha_0)\\
&\hat{y} = r_k \sin(\alpha_0)\\
&\psi'_k = \frac{\Omega_B}{2} + \theta_k - \psi_k
\end{align}
\end{subequations}
where $\psi'_k$ is conditioned to fall within $[0,2\pi]$.

Consider the angles inside the triangle, let $\alpha_1$ be $\angle{P_1P_2\rho_0}$, $\alpha_2$ be $\angle{\rho_0P_2\rho_3}$, $\alpha_3$ be $\angle{\rho_1P_2\rho_0}$, and $\alpha_4$ be $\angle{\rho_0P_2\rho_2}$. Given $l$, we also derive the quantities of the angles as follows
\begin{subequations}\label{eq:case1:alpha}
\begin{align}
\alpha_1 &= \tan^{-1}\left(\left|\frac{\hat{x}}{\hat{y}}\right|\right) \\
\alpha_2 &= \tan^{-1}\left(\left|\frac{R_B-\hat{x}}{\hat{y}}\right|\right) \\
\alpha_3&=\left\{
\begin{array}{ll}
\min\left(\cos^{-1}\left(\left|\frac{\hat{y}}{l}\right|\right),\alpha_1\right), & l>\hat{y} \\
0, & \textrm{otherwise}
\end{array}\right. \\
\alpha_4&=\left\{
\begin{array}{ll}
\min\left(\cos^{-1}\left(\left|\frac{\hat{y}}{l}\right|\right),\alpha_2\right), & l>\hat{y} \\
0, & \textrm{otherwise}
\end{array}\right.
\end{align}
\end{subequations}
and in an extreme case where $\hat{y}=0$, we simply apply $\alpha_1=\alpha_2=\frac{\pi}{2}$ and $\alpha_3=\alpha_4=0$.

The condition that the vehicle will depart from the left edge, i.e. Case 1, is
\begin{equation}
\frac{3\pi}{2}-\alpha_1 \leq \psi_k' \leq \frac{3\pi}{2}+\alpha_2
\end{equation}
and the probability that the vehicle travelling distance is shorter than $l$ is
\begin{equation}
\mathcal{P}_1(l) = \frac{\alpha_3+\alpha_4}{2\pi}.
\end{equation}

Finally, with the knowledge of the travelling direction, the departure point of the vehicle relative to beam $b_k$ is $\acute{r}_k\angle{\acute{\phi}_k}$ which can be computed by
\begin{subequations}\label{eq:case1:depart}
\begin{align}
&\acute{r}_k = \hat{x} - \hat{y} \tan\left(\frac{3\pi}{2}-\psi'_k\right)\\
&\acute{\phi}_k = \frac{\Omega_B}{2}
\end{align}
\end{subequations}
and the distance travelled $\tau_1$ is
\begin{equation}\label{eq:case1:dist}
\tau_1 = \sqrt{\hat{y}^2+(\hat{x}-\acute{r})^2}.
\end{equation}

\subsection*{Case 2: Departing From Right Edge}

For Case 2, we perform linear transformation such that $P_1$ is at the origin and the edge $\overline{P_1P_4}$ coincides on x-axis. In Fig.~\ref{fig:beamgeometry}, $\rho_3$ is $P_4$. The quantities $\alpha_0$, $\hat{x}$ and $\hat{y}$ for Case 2 are thus
\begin{subequations}
\begin{align}
&\alpha_0 = \frac{\Omega_B}{2} + \phi_k \\
&\hat{x} = r_k \cos(\alpha_0)\\
&\hat{y} = r_k \sin(\alpha_0)\\
\end{align}
\end{subequations}

Similar to Case 1, given $l$, the quantities $\alpha_1$, $\alpha_2$, $\alpha_3$ and $\alpha_4$ can be computed using \eqref{eq:case1:alpha}, and the probability that the vehicle travelling distance is shorter than $l$ can be determined by
\begin{equation}
\mathcal{P}_2(l) =  \frac{\alpha_3+\alpha_4}{2\pi}.
\end{equation}

The condition for Case 2 is
\begin{equation}
\frac{3\pi}{2}-\alpha_1 \leq \psi_k' \leq \frac{3\pi}{2}+\alpha_2
\end{equation}
where in this case %{\color{red}\sout{HL $\psi_k' -> \psi'$?}}
\begin{equation}
 \psi_k' = \frac{\Omega_B}{2} - \theta_k + \psi_k
\end{equation}
and we further condition $\psi_k'$ to ensure that its value falls within $[0,2\pi]$.

Given the vehicle travelling direction, the departure point follows the computation as shown in Case 1 given in \eqref{eq:case1:depart}. The distance travelled $\tau_2$ also uses the same computation as Case 1 given in  \eqref{eq:case1:dist}.

\subsection*{Case 3: Departing From Arc Edge}

%\footnote{\color{red}how to calculate $x_p$ given beam location and direction $(x_i,y_i,\theta_i)$? What is $x_p1$ in the arc working? \color{blue} $x_{P1}$ and $x_{P2}$ are the x-axis locations for the intersection points respectively.\color{red}look at my figure, for case 3, please express all the $\alpha$, departure point relative to the beam, distance travelled. \color{blue} Please see Eq.22, updated based on your figure. \color{red} can you provide details, $\hat{x}$, $\hat{y}$ expressions, $\rho_{1}(x)$? Make sure all details provided. Also explicit departure point relative to the beam, distance travelled please. $\mathcal{P}(l)$ is wrong, you shouldn't make this silly mistake. UPDATE: We don't need to calculate the departure point, but we still need distance travelled $\tau_3$.\color{blue} UPDATE: We now have coordinates for transformed beam in Case 3, other quantities will be obtained from \eqref{eq:case3:matrix} easily, will be added soon.}

For Case 3, we perform linear transformation such that $P_4$ is shifted to the origin and the edge $\overline{P_4P_3}$ coincides on x-axis. The transformation involves translation and rotation of the beam.

As shown in Fig.~\ref{fig:beamgeometry} for Case 3, %{\color{red}\sout{HL: Case 3}}, 
the location of the transformed $P_2$ is $(\hat{x},\hat{y})$ and the transformed direction of movement $\psi_k'$ can be computed by the following
\begin{subequations}
\begin{align}
&\hat{x} = r_k \sin(\phi_k) + R_B \sin\left(\frac{\Omega_B}{2}\right) \\
&\hat{y} = -r_k \cos(\phi_k) + R_B \cos\left(\frac{\Omega_B}{2}\right) \\
&\psi_k' = \frac{3\pi}{2} - \theta_k + \psi_k.
\end{align}
\end{subequations}

Consider the angles in the triangle $\Delta P_4P_2P_3$ shown in Fig.~\ref{fig:beamgeometry}, let $\alpha_1$ be $\angle{P_4P_2\rho_0}$, $\alpha_2$ be $\angle{\rho_0P_2P_3}$, $\alpha_3$ be $\angle{\rho_1P_2\rho_0}$, and $\alpha_4$ be $\angle{\rho_0P_2\rho_2}$. Given $l$, we can derive the quantities of the angles as follows
\begin{subequations}\label{eq:case3:alpha}
\begin{align}
\alpha_1 &= \tan^{-1}\left(\left|\frac{\hat{x}}{\hat{y}}\right|\right) \\
\alpha_2 &= \tan^{-1}\left(\left|\frac{R_B-\hat{x}}{\hat{y}}\right|\right) \\
\alpha_3&=\left\{
\begin{array}{ll}
\min\left(\sin^{-1}\left(\frac{\hat{x}-x_{\rho_{1}}}{l}\right),\alpha_1\right), & \hat{x}>x_{\rho_1} \\
0, & \textrm{otherwise}
\end{array}\right. \\
\alpha_4&=\left\{
\begin{array}{ll}
\min\left(\sin^{-1}\left(\frac{x_{\rho_{2}}-\hat{x}}{l}\right),\alpha_2\right), & x_{\rho_2}>\hat{x} \\
0, & \textrm{otherwise}
\end{array}\right.
\end{align}
\end{subequations}
where $x_{\rho_1}$ and $x_{\rho_2}$ are the x-coordinates of points $\rho_1$ and $\rho_2$ respectively. As shown in Fig.~\ref{fig:beamgeometry}, the points $\rho_1$ and $\rho_2$ are the two intersection points between the circle centered at $P_1$ with radius $R_B$ and the circle centered at $P_2$ with radius $l$. They can be determined by finding the intersection points of the two circles. In the case when $l$ is small and no intersection points can be found, we apply  $\alpha_3=\alpha_4=0$.

For completeness, we provide the solution of the intersection points of two circles in the following. Let $C_i$ be a circle centered at $(\breve{x}_i,\breve{y}_i)$ with radius $\breve{r}_i$, and here we define two circles $C_1$ and $C_2$. The intersection points of the two circles are
\begin{subequations} \nonumber
\begin{align}
&d_o = \sqrt{(\breve{x}_1-\breve{x}_2)^2+(\breve{y}_1-\breve{y}_2)^2} \\
&l_o = \frac{\breve{r}_1^2 - \breve{r}_2^2 + d_o^2}{2d_o} \\
&h_o = \sqrt{\breve{r}_1^2-l_o^2} \\
&\dot{x}_1 = \frac{l_o}{d_o}(\breve{x}_2-\breve{x}_1) + \frac{h_o}{d_o}(\breve{y}_2-\breve{y}_1) + \breve{x}_1 \\
&\dot{y}_1 = \frac{l_o}{d_o}(\breve{y}_2-\breve{y}_1) - \frac{h_o}{d_o}(\breve{x}_2-\breve{x}_1) + \breve{y}_1 \\
&\dot{x}_2 = \frac{l_o}{d_o}(\breve{x}_2-\breve{x}_1) - \frac{h_o}{d_o}(\breve{y}_2-\breve{y}_1) + \dot{x}_1 \\
&\dot{y}_2 = \frac{l_o}{d_o}(\breve{y}_2-\breve{y}_1) + \frac{h_o}{d_o}(\breve{x}_2-\breve{x}_1) + \breve{y}_1
\end{align}
\end{subequations}
where $(\dot{x}_1,\dot{y}_1)$ and $(\dot{x}_2,\dot{y}_2)$ are two intersection points.

The probability that the vehicle travelling distance is shorter than $l$ can be determined by
\begin{equation}
\mathcal{P}_3(l) =  \frac{\alpha_3+\alpha_4}{2\pi}.
\end{equation}

To derive the distance travelled within the beam for Case 3, we first formulate the departure location by using the starting location and the movement vector. The vehicle starting location in a vector form is given by $r_k\angle{\phi_k}$ and its movement vector is given by $d\angle{\psi_k}$ where $d$ is the distance travelled. By summing the two vectors, the resultant vector is pointing to the location after the vehicle travelled $d$ distance. The magnitude of the resultant vector is the distance away from the beam origin $P_1$. Since the vehicle departs the beam at the point when it is $R_B$ distance away from $P_1$, then we have
\begin{equation}
|r_k\angle{\phi_k}+d\angle{\psi_k}| = R_B.
\end{equation}
The magnitude of the resultant vector is already given in \eqref{eq:trajectory}. By setting it to $R_B$ and solving for $d$, we get
\begin{equation}
d = -r_k\cos(\gamma_k) \pm \sqrt{(r_k\cos(\gamma_k))^2-(r_k^2-R_B^2)}.
\end{equation}

The above result produces two solutions, a positive and negative values indicate forward and backward movements respective. The distance travelled $\tau_3$ for Case 3 always takes the positive solution of $d$.

\section{Derivation of $J_k^*(r_k,\phi_k,l,\psi_k)$}\label{appendix:Jstar}

We first note that $J_k^*(r_k,\phi_k,l,\psi_k)$ is a special case of $J_k(r_k,\phi_k,l)$ where the travelling direction is explicitly given. In our highway scenario, the vehicles can only move in one of the two directions. Since we orient our map such that the highway runs horizontally, vehicles can only move in the direction either $\psi_k=0$ or $\pi$ depending on which lane they are travelling on which is determined by $r_k\angle{\phi_k}$.

The calculation of $J_k^*(r_k,\phi_k,l,\psi_k)$ is straightforward. Knowing the setup of beam $b_k$, the starting location of the vehicle $r_k\angle{\phi_k}$, and the travelling direction $\psi_k$, using the working in Appendix~\ref{appendix:Jk}, we can test whether the vehicle is departing beam $b_k$ from its left, right or arc side. We can also calculate the corresponding distance travelled $\tau^*({\psi_k})$ to be either $\tau_1$ for Case 1, $\tau_2$ for Case 2, or $\tau_3$ otherwise. With the above, we have
\begin{equation}
J_k^*(r_k,\phi_k,l,\psi_k) = \left\{
\begin{array}{ll}
    1, & \tau^*({\psi_k})\le l \\
    0, & \textrm{otherwise.}
\end{array}
\right.
\end{equation}

\section*{Acknowledgment}

This work was partly sponsored by Horizon 2020 Marie Skłodowska-Curie Actions under the project SwiftV2X (grant agreement ID 101008085) and DEDICAT 6G (grant no. 101016499). The authors would like to acknowledge the support of 5GIC/6GIC at the University of Surrey.

% Can use something like this to put references on a page
% by themselves when using endfloat and the captionsoff option.
\ifCLASSOPTIONcaptionsoff
  \newpage
\fi

\iffalse

\fi

\bibliographystyle{IEEEtran}
\bibliography{ref}

% biography section
% 
% If you have an EPS/PDF photo (graphicx package needed) extra braces are
% needed around the contents of the optional argument to biography to prevent
% the LaTeX parser from getting confused when it sees the complicated
% \includegraphics command within an optional argument. (You could create
% your own custom macro containing the \includegraphics command to make things
% simpler here.)
%\begin{IEEEbiography}[{\includegraphics[width=1in,height=1.25in,clip,keepaspectratio]{mshell}}]{Michael Shell}
% or if you just want to reserve a space for a photo:

\begin{IEEEbiography}[{\includegraphics[width=1in,height=1.25in,clip,keepaspectratio]{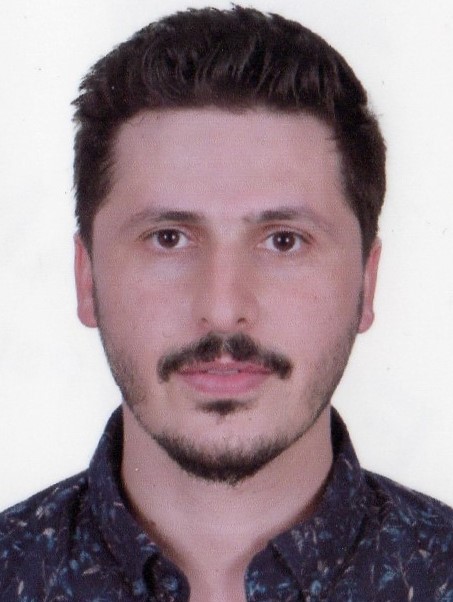}}]{Abdulkadir Kose }
received the M.Sc. degree from Yeditepe University, Turkey, in
2016,  and the Ph.D. degree from the Institute for Communication
Systems (ICS), the University of Surrey, U.K., in 2021. He is currently an Assistant Professor at Abdullah Gül University, Turkey. His recent research interests include mmWave networks, vehicular communications, mobility management, reconfigurable intelligent surface (RIS) assisted wireless communication.
\end{IEEEbiography}

\begin{IEEEbiography} [{\includegraphics[width=1in,height=1.25in,clip,keepaspectratio]{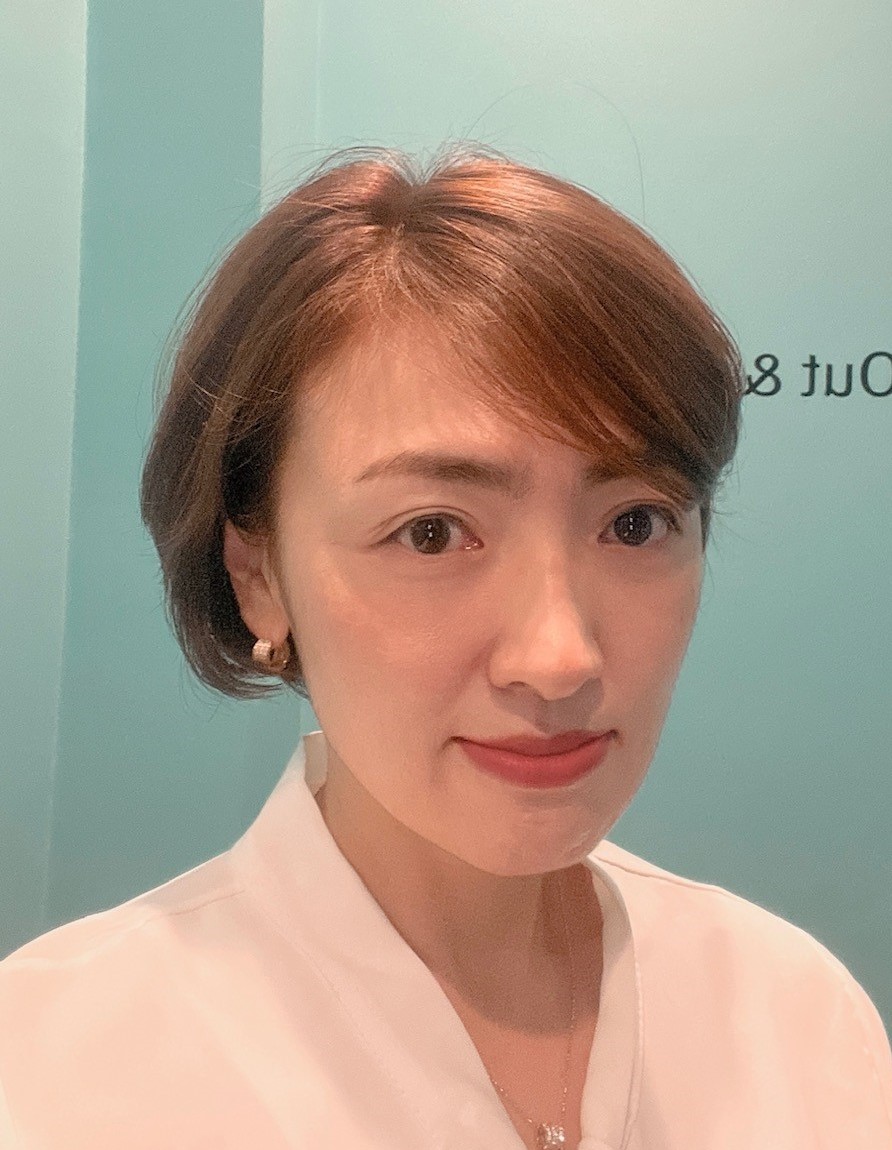}}]{Haeyoung Lee}
(Member, IEEE) earned her Ph.D. degree from the Centre for Communication Systems (CCSR) at the University of Surrey, U.K., in 2015. Between 2004 and 2007, she served as a Research Engineer at the Mobile Communication R\&D Centre, Samsung Electronics, South Korea, contributing to the development of a CDMA mobile internet platform. Following this, from 2007 to 2015, she held the position of Research Officer at the National Radio Research Agency in South Korea. In 2016, she joined the 5G Innovation Centre at the University of Surrey as a (senior) research fellow. Currently, she holds the position of Senior Lecturer in Mobile Communications at the University of Hertfordshire. Her research interests encompass radio resource management in wireless communication, optimization, machine learning, and vehicular networks. 
Since 2023, she has served as the Technical Program Committee (TPC) chair for the International Conference on Information Networking (ICOIN). Additionally, she has taken on the role of Editor for the MDPI Topic on Machine Learning in Communication Systems and Networks.

\end{IEEEbiography}

\begin{IEEEbiography}[{\includegraphics[width=1in,height=1.25in,clip,keepaspectratio]{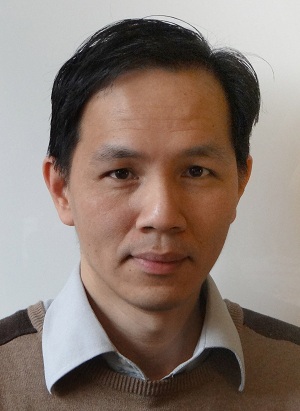}}]{Chuan Heng Foh}
(Senior Member, IEEE) received the M.Sc. degree from Monash University, Melbourne, VIC, Australia, in 1999 and the Ph.D. degree from the University of Melbourne, Melbourne, VIC, Australia, in 2002. After his Ph.D., he spent six months as a Lecturer with Monash University. In December 2002, he joined Nanyang Technological University, Singapore, as an Assistant Professor until 2012. He is currently a Senior Lecturer with the University of Surrey, Guildford, U.K. He has authored or coauthored more than 180 refereed papers in international journals and conferences. His research interests include protocol design, machine learning application and performance analysis of various computer networks including wireless local area networks, mobile ad hoc and sensor networks, vehicular networks, Internet of Things, 5G/6G networks and Open RAN. He served as a Vice Chair (Europe/Africa) for IEEE TCGCC during 2015 and 2017. He is currently the Vice-Chair of the IEEE VTS Ad Hoc Committee on Mission Critical Communications. He is on the editorial boards of several international journals.
\end{IEEEbiography}

\begin{IEEEbiography}[{\includegraphics[width=1in,height=1.25in,clip,keepaspectratio]{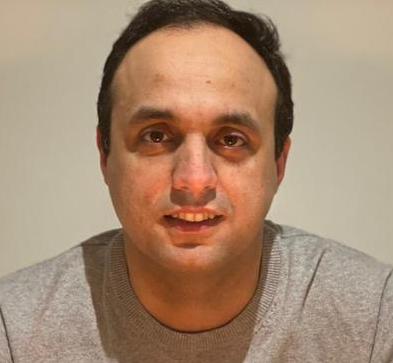}}]{Mohammad Shojafar} \textbf{(M'17-SM'19)} is a Senior Lecturer (Associate Professor) in the network security and an Intel Innovator, Professional ACM member and ACM Distinguished Speaker, a Fellow of the Higher Education Academy, and a Marie Curie Alumni, working in the 5G \& 6G Innovation Centre (5GIC \& 6GIC), Institute for Communication Systems (ICS), at the University of Surrey, UK. Before joining 5GIC \& 6GIC, he was a Senior Researcher and a Marie Curie Fellow in the SPRITZ Security and Privacy Research group at the University of Padua, Italy.  Dr Mohammad secured around £1.2M as PI in various EU/UK projects, including 5G Mode (funded by DSIT/UK;2023), TRACE-V2X (funded by EU/MSCA-SE;2023), D-Xpert (funded by I-UK;2023), AUTOTRUST (funded by ESA/EU;2021), PRISENODE (funded by EU/MSCA-IF:2019), and SDN-Sec (funded by Italian Government:2018). He was also COI of various UK/EU projects like HiPER-RAN (funded by DSIT/UK;2023), APTd5G project (funded by EPSRC/UKI-FNI:2022), ESKMARALD (funded by UK/NCSC;2022), GAUChO, S2C and SAMMClouds (funded by Italian Government;2016-2018).
He received his Ph.D. degree in ICT from Sapienza University of Rome, Rome, Italy, in 2016 with an ``Excellent'' degree. He is an Associate Editor in \textit{IEEE Transactions on Network and Service Management}, \textit{IEEE Transactions on Intelligent Transportation Systems}, \textit{IEEE Transactions on Circuits and Systems for Video Technology} and Computer Networks. For additional information: \url{http://mshojafar.com}
\end{IEEEbiography}

% insert where needed to balance the two columns on the last page with
% biographies
%\newpage

% You can push biographies down or up by placing
% a \vfill before or after them. The appropriate
% use of \vfill depends on what kind of text is
% on the last page and whether or not the columns
% are being equalized.

%\vfill

% Can be used to pull up biographies so that the bottom of the last one
% is flush with the other column.
%\enlargethispage{-5in}

% that's all folks
\end{document}